# Decentralized Resource Discovery and Management for Future Manycore Systems

Javad Zarrin[1] (jz461@cam.ac.uk),
Rui L. Aguiar[2] (ruilaa@ua.pt),
Joao Paulo Barraca[3] (jpbarraca@ua.pt)

## Abstract

The next generation of many-core enabled large-scale computing systems relies on thousands of billions of heterogeneous processing cores connected to form a single computing unit. In such large-scale computing environments, resource management is one of the most challenging, and complex issues for efficient resource sharing and utilization, particularly as we move toward Future ManyCore Systems (FMCS). This work proposes a novel resource management scheme for future peta-scale many-core-enabled computing systems, based on hybrid adaptive resource discovery, called ElCore. The proposed architecture contains a set of modules which will dynamically be instantiated on the nodes in the distributed system on demand. Our approach provides flexibility to allocate the required set of resources for various types of processes/applications. It can also be considered as a generic solution (with respect to the general requirements of large scale computing environments) which brings a set of interesting features (such as auto-scaling, multitenancy, multi-dimensional mapping, etc,.) to facilitate its easy adaptation to any distributed technology (such as SOA, Grid and HPC many-core). The achieved evaluation results assured the significant scalability and the high-quality resource mapping of the proposed resource discovery and management over highly heterogeneous, hierarchical and dynamic computing environments with respect to several scalability and efficiency aspects while supporting flexible and complex queries with guaranteed discovery results accuracy. The simulation results prove that, using our approach, the mapping between processes and resources can be done with high level of accuracy which potentially leads to a significant enhancement in the overall system performance.

## 1. Introduction

Modern large-scale distributed computing systems are undergoing with the rapid evolution of processor and network architectures. And they have made possible: i) the integration of more and more cores into one single chip; ii) many-chips being interconnected into a single machine; iii) more and more machines getting connected with highly increasing bandwidth. This leads to the emergence of the next generation of manycore enabled large-scale computing systems which rely on thousands of billions of heterogeneous processing cores connected to form a single computing unit. Current large-scale computing environments such as HPC clusters (e.g., Infiniband-based distributed memory machines, Bewolf clusters), Grids and Clouds are common scenarios when discussing enhancements to overall computing/system performance and resource/data/service/application accessibility through efficient sharing and utilization of the integrated infrastructures and hardware resources (such as computing, storage, data and network resources) in large-scale systems with high heterogeneity (in terms of resources, applications, platforms, users, virtual organization, administration policies, etc.) and high dynamicity. In such large-scale computing environments, resource management is one of the most challenging, and complex issues for efficient resource sharing and utilization, particularly as we move toward Future ManyCore Systems (FMCS).

---

1 Javad Zarrin from Computer Laboratory, University of Cambridge, UK
2 Rui L. Aguiar from Universidade de Aveiro and Instituto de Telecomunicacoes, Aveiro, Portugal
3 Joao Paulo Barraca from Universidade de Aveiro and Instituto de Telecomunicacoes, Aveiro, Portugal

There are various types of techniques and methods to control and manage the infrastructure for each one of the aforementioned computing environments, which differ based on their main focus, embedded technologies and system architectures. And in fact, designing a resource management architecture which can be applied and adjusted to the requirements of these different computing environments is an extra challenge.

In this report, by "manycore enabled computing systems", we mean future computing systems that support thousands of chips per compute node and thousands of heterogeneous cores per chip (as predicted in [1]). We address the problem of resource management for this future large-scale manycore enabled computing environment. In such large-scale systems (e.g., future multitenant Clouds, petascale manycores, and heterogeneous clusters) it is not feasible for the control system to centrally and statically have a complete and perfect knowledge of the entire system due to the magnitude and diversity of the amount of cores/processors and other resources. For uniformity of discussion, we will address all these control systems as distributed operation systems (DOS), although their realization may be quite different, depending on the specific large-scale computing technology under discussion. Furthermore, the variety of new emergent applications and the heterogeneous resources distributed across the network owned by different organizations and administrative domains, demand that these DOSs support an efficient mechanism for both process and resource management. Designing a DOS resource management system that meets the requirements of such large-scale environments is challenging due to several issues: a) Supporting adaptability, scalability and extensibility, b) Incorporating computing resources within different virtual organizations (VOs) while preserving the site autonomy (i.e., the distributed ownership of the resources), c) Co-allocating resources, d) Supporting low-cost computation, e) Supporting Quality of Service (QoS) guarantees.

This report introduces ElCore, a new elastic, scalable, accurate and dynamic resource management architecture for distributed systems in future large-scale manycore enabled computing environments, which supports high quality resource mapping and allocation (i.e., resource allocation accuracy) in a fully decentralized manner. In our work, we mainly focus on scalability and accuracy issues, since in our target computing environments (future large-scale manycore system), one of the most important requirements for resource management is the ability to deal with very large number of heterogeneous resources.

ElCore is based on HARD3 which is a very fine-grained, manycore level, resource discovery protocol for fully decentralized distributed systems.

Using ElCore, applications/processes can be distributed among a set of automatically self-organized and self-configured Resource Manager (RM) Entities or Resource Manager Components (RMCs) in the system. RMCs can individually control and monitor their own collection of resources while resources can dynamically be traded (transacted) among RMCs on demand basis. This is achieved by the usage of HARD3, as an embedded dynamic resource discovery method which is able to efficiently discover the most accurate and appropriate resources with the lowest possible latency (i.e., nearest matching resources) among many available resources belonging to different RM entities.

ElCore provides scalability since its decentralized discovery mechanism (i.e. HARD3) allows scalable communications and resource trading between RMC instances. In fact, HARD3 supports a highly sophisticated query and resource description model (from the very low level of processors and processes to the level of computer clusters with respect to inter-resource and inter-process communication constraints). Using HARD3, ElCore can enable the requesters to accurately discover the resources based on the required specifications and constraints for each given query. These features make ElCore flexible to be efficiently adapted to the requirements of future large-scale many-core enabled systems, be the future HPC clusters, Grids or Clouds.

ElCore also supports elasticity (roughly defined as the ability to address variable run-time changes on resources and workloads) with respect to the following two aspects. First, in our approach, we employ several resource management entities which are distributed across the system in a fully decentralized fashion. Each RMC instance can flexibly control and modify its own pool of resources according to on-demand basis. This gives elasticity to each instance of RMC to dynamically meet various, arbitrary scale users' resource requirements, ranged from small applications to large computing-intensive tasks (we discuss ElCore policy to trade resources among different instances of RMC in Section 4.7.3). Second, similar to the requirements of the current Cloud computing technology [2], we can expect that one of the most important requirement for application execution in future large-scale manycore enabled systems would be the ability of the system to deal with dynamic applications which have high variability fluctuations in their workloads. As an example of this type of applications we can mention malleable applications [3] where malleability is defined as the capability of the application to dynamically modify its data size and also number of computational entities (threads). The dynamicity of applications (in term of dynamic workload) affects the performance of resource management systems, since resource mapping is based on initial QoS requirements of applications, when workload changes (during run-time), the application execution might fails to meet its initial QoS requirements. ElCore provides elasticity to dynamically detect overloaded resources and reallocate additional resources, allowing migration of processes (like new threads) from overloaded to newly assigned resources (see Section 4.7.5). Furthermore, the modularity and Service-Oriented Architecture (SOA-based) design of ElCore allow it to be coherently integrated and used within DOSs.

In this report, for better illustration and evaluation of our approach, we use the term "resource" to refer to "computational resource" and we discuss resource management mostly from the point of view of computational capacity. Nevertheless, our resource management model is fully generic, and applicable to other types of resources (e.g., storage and networking resources).

The remainder of report is organized as follows: In Section II we discuss resource management challenges for large scale systems. Section III elaborates on some important requirements for designing a resource management model for future large-scale manycore systems. In Section IV we explain our proposed resource management architecture (DOS level discussion). In Section V, we discuss our approach with respect to FMCS. Section VI provides detail on our manycore simulation methods and presents the simulation results. Finally, Section VII presents our conclusion.

## 2. RM Challenges for Large scale Systems

In this section, we investigate resource management issues for large scale computing environments such as HPC, Grid and Cloud in the matter of future manycore architectures.

### 2.1 High Performance Computing Systems

High Performance Computing (HPC) is generally recognized as a role model for manycore systems due to the essential similarity of their architectures and hence the means of program development for both of them (HPC and manycore systems). At the same time, the manycore movement leads to a sudden boost of scale in high performance computing without essentially increasing the requirements or costs. Many HPC cluster systems employ a centralized resource allocation and management, where the jobs are being controlled centrally and statically by a single resource manager that has complete knowledge of the system. In fact, HPC clusters are generally based on a batch scheduler and resource management system which allows users to submit their jobs to a batch system where a central scheduler and resource manager entity that is hosted on the cluster-head makes decision (in terms of timing aspects and the physical location of resources) for the resource allocation to run a given job. A centralized resource manager generally manages a pool of resources and runs the jobs based on a prioritization policy. However, it may distribute the jobs among a set of other resource manager entities in a hierarchical fashion. The resources might be organized in single or multiple queues (aka partitions) based on some common characteristics of resources. Slurm [4, 5] TORQUE [6], LSF [7], PBS [8], Load Leveler [9], CCS [10] and Univa's Grid Engine (formerly Sun Grid Engine) [11] are some of the popular centralized schedulers and resource managers for the HPC clusters. HTCondor [12, 13] is another example of centralized resource management in High Throughput Computing (HTC) environments which is able to utilize large heterogeneous clusters where large numbers of relatively small-sized and short-live jobs are processed.

The future manycore enabled HPC clusters cannot rely on the centralized and static resource management approaches. Resource management for large-scale manycore systems is inherently a NP-complete problem. The large number of heterogeneous cores and applications with various requirements causes scalability issues for centrally acting heuristics, where a centralized component must maintain a global view of the entire system. Resource management itself can become a bottleneck due to high computational requests and communication latencies. Furthermore, the amount of the detailed information on the resource characteristics which is maintained by the central resource managers would be limited to a very abstract level and this leads to reducing the accuracy of resource mapping and allocation (i.e., the quality of resource mapping) for the applications. On the other hand, managing dynamic resources (when resources join, leave or fail) with dynamic attributes and behaviors becomes more complex for the central resource managers since the solutions such as periodical or triggered updating, for the large number of resources and state changes, impose significant communication overheads on the system.

In decentralized resource management, multiple resource management entities cooperate to keep the workload for all computing resources in different pools or clusters balanced and satisfy the user requirements. REXEC [14], Tycoon [15] and Cluster-on-demand [16] are some examples of decentralized resource managers for HPC clusters.

## 2.2 Grid Computing Systems

In Grid computing where multiple geographically distributed and heterogeneous clusters are combined into a single system, according to the type of Grid, there exists different types of resource management system particularly designed for computing Grids such as Nimrod$/$G [17, 18], 2K [19] (on-demand hierarchical), Bond [20] (on-demand flat), Condor [21] (computational flat), Darwin [22] (multimedia hierarchical), Globus [23] (various hierarchical cell), Legion [24] and Ninf [25] (computational hierarchical). Similar to HPC clusters, it is also common to use centralized resource management in Grid computing where users could submit their jobs to a single resource management or scheduling entity that makes decision on which cluster and nodes run the job. Grids also deploy decentralized approaches for managing resources in the system.

However, there are some major drawbacks with the current decentralized resource management schemes in HPC clusters and Grids. Most of these approaches do not consider the dynamicity of the applications (e.g., the application load might change during the run-time) and the resources (e.g., resources might join, leave or fail) in computing environments and they statically pre-assign a group of resources to each resource manager (RM) entity in order to provide a load balanced solution to distribute jobs among different queues or RM entities in the system. This limits the dynamicity of resource management, since the resources are not allowed to be traded among different RM entities in the system (i.e., each resource is managed by a fixed RM entity). In other words, when a job is submitted to a RM, if the resources in RM resource-pool do not satisfy the query conditions the RM is not able to dynamically borrow resources from other RMs in the system and this results to either failing the query or forwarding the query to other resource manager entities in the system. Efficient decision making to forward queries among the distributed RMs leads to a resource discovery problem. Some of the decentralized RM approaches employ resource discovery techniques to manage query-forwarding between RMs, but these discovery mechanisms generally are not really dynamic and decentralized. For example NimrodG [17, 18] employs MDS resource discovery [26] which is based on publish/subscribe method to maintain resources in a set of hierarchical LDAP directories. MDS hierarchical discovery approach still suffers from single point failure mentioned for centralized methods because at each level there is a central data base server responsible for resource update requests. Moreover, MDS does not scale well in computing environments with frequent updates and large numbers of application requests.

## 2.3 Cloud Computing Systems

In Cloud computing systems, resource provisioning using on-demand virtualization and VM migration technologies enable the data centers to consolidate their computing services and use minimal number of physical servers. In fact, the virtualization allows workloads to be deployed and scaled-out quickly through the rapid provisioning of Virtual Machines (VMs) on physical machines. But similar to HPC cluster and Grid, Cloud computing systems still need to deal with resource allocation and resource discovery problem to efficiently map the virtual machines to the real physical resources.

An efficient resource management (in IaaS Cloud service model) can potentially provide significant number of benefits specifically in terms of quality of service, scalability, throughput and utility optimization, cost efficiency and overhead/latency reduction. However, for future diverse, large-scale systems, there are issues such as resource modeling (description), estimation (identification), provisioning, discovery, brokering, mapping, allocation and adaptation that it should yet deal with.

Implicitly, the scale in Cloud systems must be regarded as horizontal where instances of data and code are replicated to increase availability. Whilst this is particularly useful for data and service hosts (eBay, Amazon, etc.), it does have little or no impact on scalable systems acting on vertical scale (i.e., where the actual application is split into multiple processes or threads that jointly contribute to the application's functionality). Recently there have been more and more references to the so-called HPC Cloud [27]. However these either focus on making small parallel machines (for instance, in the order of 8 cores in Amazon EC), or provide access to thousands of cores in an almost unconnected fashion, similar to Compute Grids, (like Plura Processing). In the current literature there are several types of resource management systems for Cloud computing. For instance we can mention OpenNebula [28], Eucalyptus [29], In-VIGO [30] and Cluster-on-Demand [31].

## 2.4 Manycore Systems

Current approaches for resource management in manycore systems can be categorized to off-line, mixed and on-line approaches. The off-line approaches such as [32-35] are based on priori (static) knowledge about the whole system states and resources, thus they cannot resolve the unpredicted dynamic situations particularly when the application workload dynamically varies at run-time. The mixed approaches such as [36-38] are based on the pre-calculation of all potential application mapping and selection of the best mapping at run-time. This also requires that all input applications and all possible combinations of mapping (for those applications) must be known at design-time. These approaches generally employ a central entity to select the best mapping amongst the pre-calculated solutions at run-time. But the application behavior and sequence of execution might be changed at run-time. Dynamic workload is not supported by mixed approaches. In online mapping [39-42], instead of pre-determination, the application mapping is dynamically decided at run-time. However we must note that online mapping trades off the resource allocation accuracy (i.e., the quality of resource mapping) versus computational complexity and scalability.

# 3. Main Requirements

This section provides some key aspects required to design an efficient resource management model with respect to the requirements of future large-scale manycore systems. It also must be taken into account that our proposed approach does not address all the shortcomings of prior works. Nevertheless, we address the following key aspects for future DOSs:

**System Size:** In the future, billions of devices may be connected to form a single computing system. This means that, in order to still be able to efficiently use such computing environments, we need to provide scalable approaches which can deal with so very large-scale systems. Thus, scalability becomes one of the most important requirements for resource management for future computing systems [43, 44]. Along this line, decentralized models are generally the approaches that lead to maximum scalability [45], since centralized models commonly suffer from several inherent drawbacks (such as single point of failure and communication bottlenecks). On the other hand, implementing purely decentralized resource management models might hugely increase the level of system complexity (and in some cases, it may not be feasible). Distributed hierarchies (which used for our work) can be considered as a realistic alternative decentralized model for resource management, since it can provide scalability while avoiding to overly increase the system complexity.

**Heterogeneity:** For future massively parallel, distributed and heterogeneous systems, capability of resource management approaches to efficiently handle high heterogeneity of resources is an essential requirement [46]. High heterogeneity of resources (i.e., core diversity) results in increasing the complexity of organizing and managing the resources. For example, different resources, positioned in the same chip or node, may provide different types and levels of capabilities which makes it difficult to apply a common resource management policy to efficiently explore all those heterogeneous resources. In order to solve this problem, we introduce the concept of virtual node (vnode) where each group of homogeneous resources, positioned in a common vicinity, participates to form a single vnode based on given vnode clustering parameters (we discuss this further in Section 4.7.5). In addition, we must note that using a shared memory programming or communication model is not feasible for heterogeneous cores, as well as affecting the scalability of the system. Thus, we assume message passing for communication between cores and chips.

**Mapping:** For future scenarios, with very large numbers of diverse applications and resources, high quality process-resource mapping can be identified as one of the most urgent problems to be solved [47–49]. The quality of mapping is directly relevant to the accuracy of both resource identification (i.e., precise identification of required resources for each given process) and resource allocation (precise process-resource mapping to meet process requirements). In our work, assuming the former can be dynamically performed by process managers (we discuss this further in Section 5.5), we focus on the later. In order to achieve high resource allocation accuracy, we provide a matching strategy which uses a decentralized strategy (based on distributed hierarchies) to find and map resources. Our mapping strategy supports querying for both design-time (static workload) and run-time (dynamic workload) application requirements through enabling dynamic resource discovery and facilitating process migrations (for overloaded resources). We further discuss our matching strategy in Section 4.7.5.

**Elasticity:** The elasticity can be defined as the capability of the resource management system to flexibly and sensitively behave on demand basis where the demand comes from resource manager users (requests for resources). Elasticity is one of the desired requirements for resource management in current Cloud computing systems [50–54]. It must also be considered as an important resource management requirement for future manycore systems, due to the trend of moving Cloud computing concepts to manycore systems. As we mentioned earlier, ElCore provides elasticity with respect to the amount of resources controlled by each resource management entity and the amount of resources assigned to each process.

## 4. Resource Management Architecture

We propose a novel generic resource management architecture for DOSs, which is able to efficiently manage the allocation of resources in large-scale computing environments with high rate of dynamicity and heterogeneity of resources. We assume a reliable (no message loss, duplication or corruption) and asynchronous message passing model, for inter-core and inter-chip communication in such environments. Our system covers all levels of resources, in all different levels of clustering.

We organize the distributed resources in the computing environment (containing a large number of interconnected resources in different levels) in a set of distributed hierarchies through leveraging our dynamic self-organization and self-configuration approach [54, 55]. We define the entities of Main-Control (i.e., DOS main-kernel) and Nano-Control (i.e., DOS nano-kernel) which are providing maximal and minimal numbers of capabilities, functionalities and services in the system through running each kernel on multiple cores [56-58]. These control entities differ in terms of service types, which they can dynamically instantiate on demand. The instances of these entities are positioned in the system in a way to map the structure of the underlying distributed hierarchies (i.e., deploying the main-control instances in the hierarchies' head-nodes and the nano-control instances in the leaf-nodes).

Similar to the Multitenant Clouds architecture, each (main/nano) control instance provides a dedicated share of the instance including a set of dynamic on demanded services to client machines/components in their region in the hierarchy. In fact, this approach could potentially enhance the multitenancy model for resource management to work in the hierarchies and support dynamic service instantiation (although we do not highlight this aspect since it is out of scope for this dissertation). Moreover, we classify control services as atomic-services (e.g., for the resource management service where every island of distributed resources in the system is managed by a single instance) and non-atomic services. For the non-atomic-services, for instance, we can consider the scheduling service: a local scheduler instance per tile / processor may be responsible for scheduling all code segments within that domain, but will itself be subject to a higher-level scheduler which triggers the respective code block as a whole. The atomic-services are the ones that should not be subdivided in terms of functionalities and workloads.

Figure 1 highlights the key services associated to control decisions for resource management (aspects such as deployment instantiation on process movement are important but are not involved in the resource management decisions). These services include:

**RR or Resource Requester** is a client-side discovery component, which is responsible for generating, starting and disseminating a query in the system finally returning information about the requested resources (e.g., available qualified processing cores and their respective capabilities). It is an atomic service that can be instantiated by any control entities. Each potential resource client in the system can also have a single RR instance.

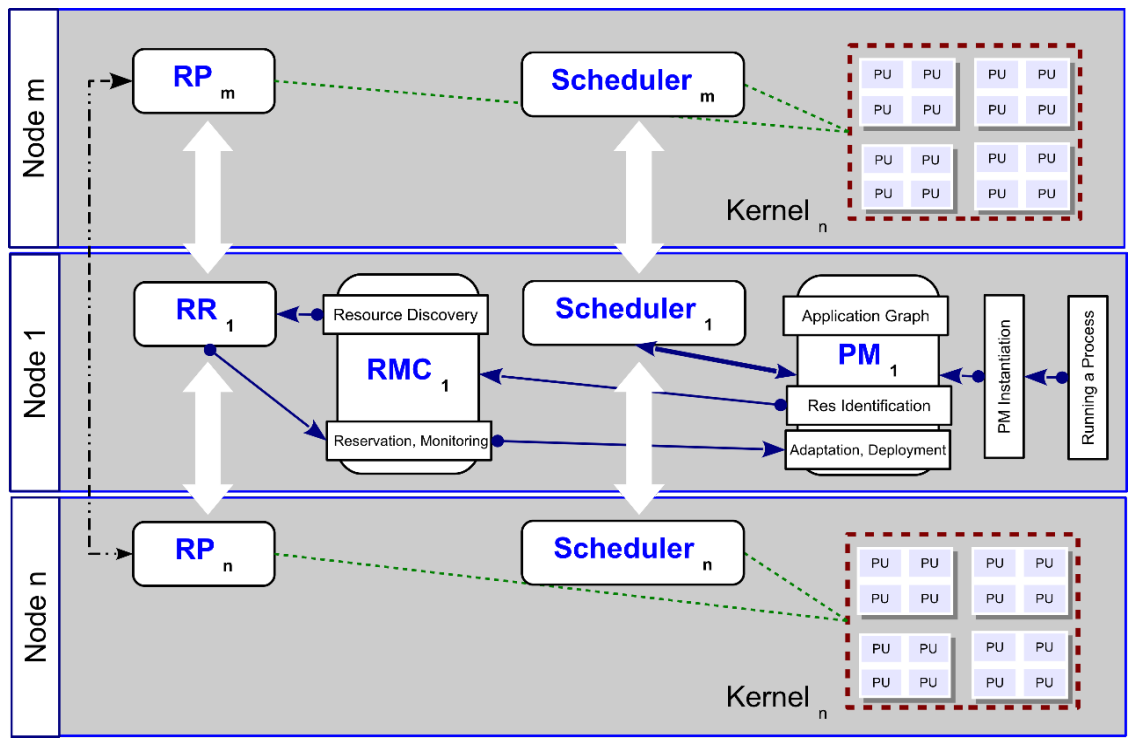

Figure 1: Resource management architecture.

**RP or Resource (Information) Provider** is a server-side discovery component, which directly offers information services to Resource Requesters and communicates with other Resource Providers in the distributed system. RPs collaborate with each other to resolve the received queries from RRs in a completely decentralized and distributed fashion. It is an atomic service which can only be instantiated by main-control instances. However, depending on the specific level of hierarchy, it can be configured as a non-atomic service considering a wider searching space. The RR-RP model is based on self-organizing processing resources in the system, where the computing resources are organized into distributed hierarchies according to a hierarchical resource description (i.e., multi-layered resource description). Moreover, in each layer different algorithms and adapted mechanisms (such as distributed hash tables, distributed probability tables and any-casting) are implemented in order to redirect the discovery requests to the proper requested resources within the layers (e.g., leaf-node, aggregate-node and supernode layers) in the system. More details of our RR-RP resource discovery model (called HARD) is presented in Section 5.

**Scheduler** is a non-atomic service (since the scheduling loads can be splitted among the other scheduling entities), which is in charge of executing the threads according to a certain order. In general, both types of control entities instances are able to instantiate scheduler instances on demand (e.g., in cluster computing environment, there will be one such component per each processor/resource where the application threads are loaded and needed to be executed.). The details of the scheduling mechanism used are presented in [60].

**Process Manager (PM)** is in charge of running an application. One such component is loaded and starts every-time the user issues a command for starting an application (i.e., one PM per user per application). The PM is an atomic-service and there would be one single instance for each application. This facilitates the usage of our

## 5. Resource Discovery Architecture

This section discusses the system architecture for our proposed 3-layered Hybrid Adaptive Resource Discovery (HARD3) Protocol. HARD3 provides a RR-RP discovery model which is used for information exchange between resource management components distributed in the system.

Modern and future distributed computing systems are highly-hierarchical and highly-heterogeneous in nature. Accordingly, the HARD3 architecture deploys various layer-based hybrid adaptive mechanisms (i.e., interlayers and intra-layers methods) in order to efficiently direct discovery requests to the proper resources across layers. This means that, according to properties and characteristics of each layer in the hierarchy, HARD3 proposes a set of specific adapted methods, which have been designed to obtain the maximum discovery efficiency on the target layer, while an integrated and coherent approach is used to traverse layers in hierarchy. Figure 2 depicts the overall architecture of HARD3, highlighting users, main services, underlying techniques and organization of computing resources in different layers.

We build our discovery system architecture based on a self-organized virtual hybrid overlay. In order to create the virtual overlay, at first, the unstructured resources in the system are organized within vnodes according to their homogeneity and proximity parameters (i.e., their similarities and locations). In the next step vnodes start to negotiate with each other in a multi-round distributed fashion to seek agreement on the contribution (i.e., module-role or vnode type) of each party in the overlay hierarchy. As negotiations evolve, each vnode shapes its own system-view by improving and consolidating its own knowledge on the entire system. The resulting overlay contains three different types of virtual-nodes: Leaf-Nodes (LNs), Aggregate-Nodes (ANs) and Super-Nodes (SNs) which take position in layer-ln, layer-an and layer-sn of the hierarchy respectively. Depending on the vnode type (i.e., module-role), each virtual-node provides different HARD3 services (e.g., RP-QMS and RP-SQMS). vnodes in the upper layers are able to provide discovery services specific to their own layer and all the services in the lower layers. vnodes respond to the discovery demands based on their module-roles as well as the immediate requirements of the triggered communication events. For example, a vnode in layer-sn (i.e., a super-node) provides RP-SQMS service. However, depending on the properties of the received communication events, it may also provide RP-QMS or RR services or participate in the overall discovery procedure by playing a role of a leaf-node.

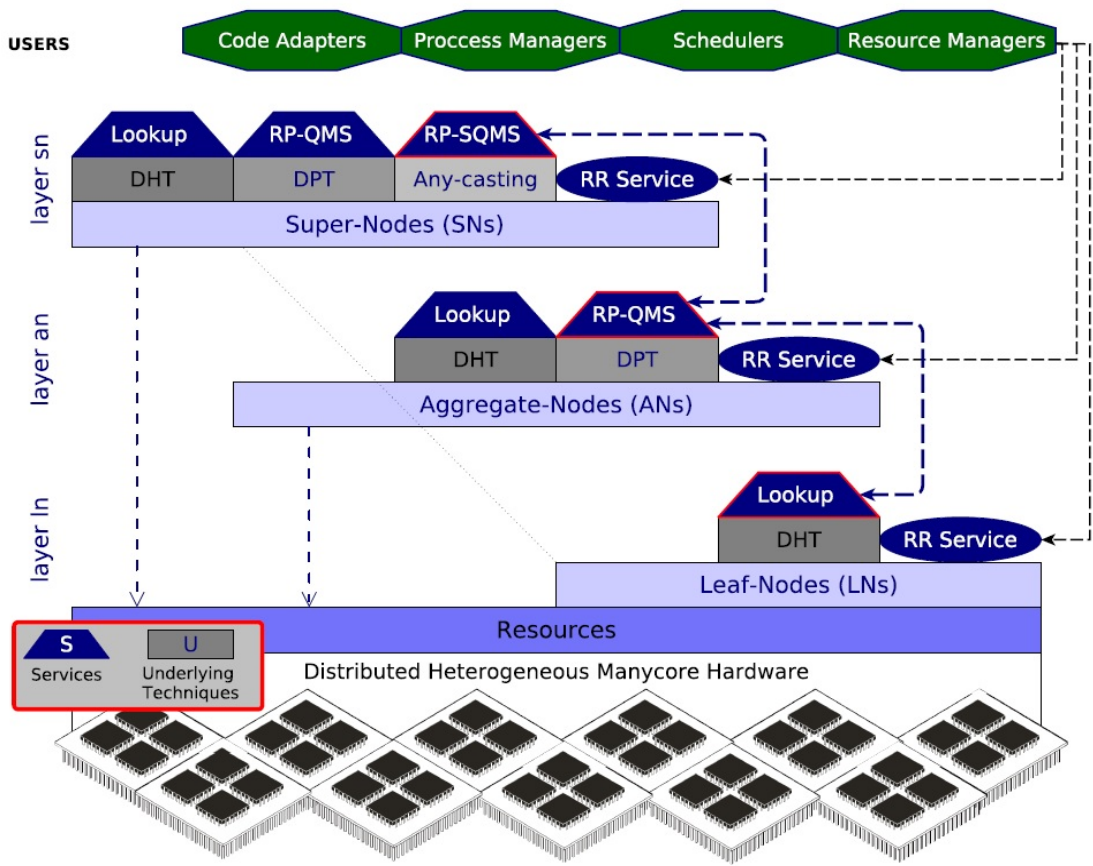

Figure 2: Resource Discovery System Architecture.

As it is represented in Figure 2, the leaf-nodes in layer-ln are organized in Distributed Hash Tables (DHTs) based on the multidimensional fingerprint (layer-stamp) of each participating vnode. In fact, the leaf-nodes participate in a core-level specification-based DHT ring where the sibling nodes (i.e., the vnodes with similar resources) are linearly organized in linked lists with single entries on the DHT ring. Similarly, aggregate-nodes (in layer-an) and super-nodes (in layer-sn) regardless of their module-role, participate in DHT. For each group of LNs which elect a single AN/SN as their common resource provider (RP-QMS or RP-SQMS), the DHT is composed of all the vnodes in the group (containing LN members and either a single AN or a single SN). In other words, all vnodes, regardless of their module-role, are able to perform Lookup queries over DHTs.

The reason to use DHT in the core-level (layer-ln) of our architecture is due to the following aspects: (a) DHT is scalable: this is especially important when considering the potentially large number of cores that can reside on a single die (in future systems). (b) DHT is fast, reliable, fault tolerant and deterministic: resource discovery in the core-level is much more sensitive to speed than querying in the network-level, due to the tightly coupled design of many-core processors. In many-core level, resource discovery might be ineffective if it fails to provide required information in an adequate amount of time (e.g., discovery latency might have a direct impact on the cost of execution migration in a many-core environment). Moreover, in such highly sensitive environments, it is essential for a discovery method to operate reliably and provides deterministic results (undetermined results might have cost by reprocessing the query or exploring an already visited search space). (c) DHT maintenance is low cost (in terms of memory and communication): a very small finger-table is required to be maintained in each vnode in layer-sn. (d) DHT supports attribute-based query description: this makes DHTs more compatible to our attribute-based resource description model. On the other hand, DHTs originally do not support semantic-based querying. To solve this issue, in our DHT variation, we enhanced the original Chord DHT to support a similarity algorithm which makes feasible similar-matching instead of exact-matching (HARD3 supports both modes of matching through specifying the desired matching mode in the query by the user).

Query Management Service (QMS) is a service which provides query processing facilities in layer-an. It uses a probability-based mechanism to guide queries among a group of aggregate nodes which share a single super-node as the resource provider (i.e., RP-SQMS). During the discovery procedure, Distributed Probability Tables (DPTs) cooperate with each other in a set of dynamic distributed learning processes, which are adapted to the progressive environmental changes. For each AN, its local probability table dynamically collects, aggregates and updates information about the status of the overall resources in the system, gathered from all transacted queries and results through the AN itself. By using this DPT technique, the network that connects ANs becomes increasingly resource-aware, as the number of traversed queries increases across the system.

We use DPT as a base method in the die-level (layer-an) due to its scalability, dynamicity, efficiency and also its compact structure. Comparing to DHT, DPT provides probabilistic results instead of deterministic results. But this not a drawback, since DPTs operate in the middle-level of HARD architecture which does not need to provide deterministic results. The reason is that, queries are not going to be concluded in layer-an. In fact, a query processing starts from the top-level (layer-sn) (of course, if there exist any query conditions for this layer) and then goes to the middle-level (layer-an) and finally it could be concluded in the lower-level (layer-ln). Furthermore, we enhance our DPT approach by introducing a SoR mechanism which can help DPT to provide deterministic results whenever feasible.

The compact design and dynamic nature of DPT provides a facility to efficiently cope with dynamic changes in the environment (e.g., unavailability of resources due to resource failure, resource reservation, etc). Each vnode in layer-an maintains a small probability table. Depending on the number of predefined attributes in the system, a property table may include multiple records (called resource-type or resource-category records), where each record represents the aggregated probability information for all neighbors with respect to the overall query transaction data (monitoring data), collected and analyzed, for a single attribute over a predefined specific range of values. In fact, each resource-type record includes probability factors for all neighbors as well as a suggestion of a SoR (a vnode which deterministically can provide resources, matched with the resource-type definition of the record). We also note that probability tables only cover attributes defined for layer-ln and layer-an.

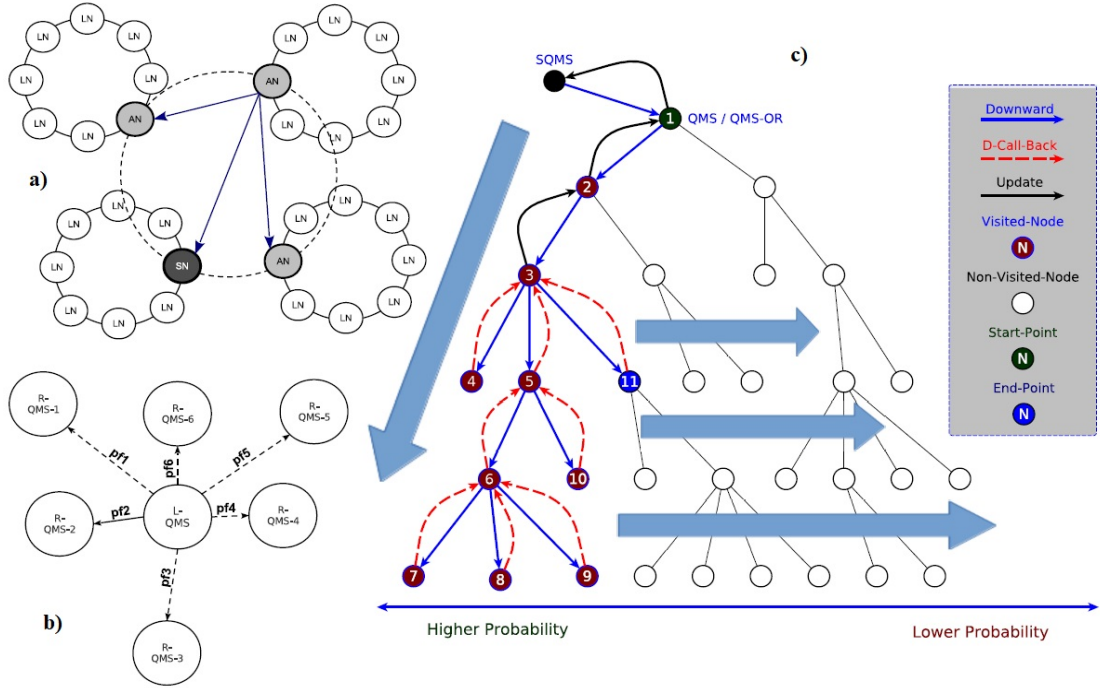

Figure 3: a) Types of vnodes (SNs, ANs, and LNs) in HARD3 hierarchical structure and their relevant virtual clustering. b) Selection of the next QMS based on values of probe factors and probabilities. c) Example of DPT's tree exploration in HARD3.

As it is shown in the Figure 3-a, vnodes are self-organized in virtual groups, as SNs, ANs and LNs groups in a hierarchical manner. Communications between vnodes are either within layer or across layers. In Figure 3-b, a vnode of type AN analyzes the probability to find the best match for a given query among a set of its neighboring ANs. Figure 3 demonstrates an example of DPT tree exploration. ANs in layer-an are automatically organized in a DPT tree. A Depth First Search (DFS) mechanism forms the base structure to explore aggregate-nodes within the layer-an. DPT and other search polices for selecting an AN among the neighbors are applied on top of the DFS tree.

Super Query Management Service (SQMS) is a specific QMS which provides additional capabilities to support query forwarding in layer-sn. For instance, as we can see in Figure 2, super-nodes (i.e., SQMS providers) are able to concurrently provide multiple services (such as lookup, QMS, SQMS and RR) for the different triggered communication events. The query forwarding in layer-sn uses the specifications of the resources in the node-level to conduct a specification based anycasting method to direct the queries among SNs. It uses the top level layer-stamps to create anycast groups, while nodes in this layer are able to automatically adjust to the anycast group they are interested in based on their specifications in layer-sn. We use anycasting as a base method for querying in the network-level (layer-sn) due to its scalabity, efficiency and its powerful features which make it more adequate and compatible for resource discovery in computing systems with a large number of network connected nodes.

Depending on the specific DOS architecture, HARD3 users could be resource management entities, schedulers, process managers or even code adapters. Each user would be able to perform resource discovery through invocation of a RR entity. RR in turn sends the given query to its local RP-QMS. Due to the type of query and the user's demand (e.g., simple single resource, multiple heterogeneous resources, complex resource graph containing the constraints for inter resource communications, etc.) RP-QMS splits the main-query to a set of sub-queries and chooses the appropriate layer that each sub-query must start to process. Finally, the RP-QMS that originated the sub-queries, aggregates the discovery results, and responds the RR with a set of resource matches that optimally satisfy the main-query's demand.

## 6. Resource Management Component

Following this section, we discuss the RMC mechanisms in more detail. RMC manages, monitors and controls the resources in the distributed system. It keeps, monitors and modifies the status of the resources, which are used by the local process managers. It invokes resource discovery components (i.e., RRs and RPs) to find the list of required resources according to the incoming request (wishlist) from Process Managers (PMs) or any other DOS components. It is also able to reserve a single resource or set of resources for a specific process/thread/threads, in order for the result of resource discovery (originated by other remote instances of RRs in the distributed system) not to contain the occupied resources. The reserved resources would be released upon receiving the notification (i.e., notification of process execution completion) from PMs. Additionally, RMC provides a set of capabilities to manage and monitor resources for the other DOS components like the Process Manager. In order to allocate the proper available resources to the application the Process Manager needs to query the Resource Manager.

RMC provides support for distributed management and control of resources. Every pool of resources is managed by one RMC instance, and each instance can provide a global view of all potential resources in the network. It receives requests from many PM components, while each request will specify a set of resource requirements (see Figure 4). For example a PM may ask for a minimum and maximum number of processors per each kind, for a certain fraction of the network bandwidth. RMC checks if the required amount of resources is available, and in case of a positive answer it returns a set of resource descriptors that the PM caller module can use to run its tasks. The interaction between the PM and the RMC can be very simple (one simple request that can be accepted or rejected) or very complex (based on a negotiation protocol), depending on the specific DOS configuration.

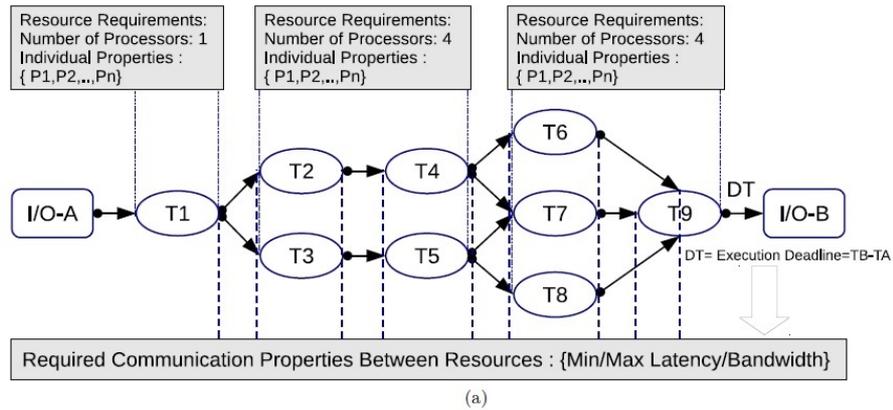

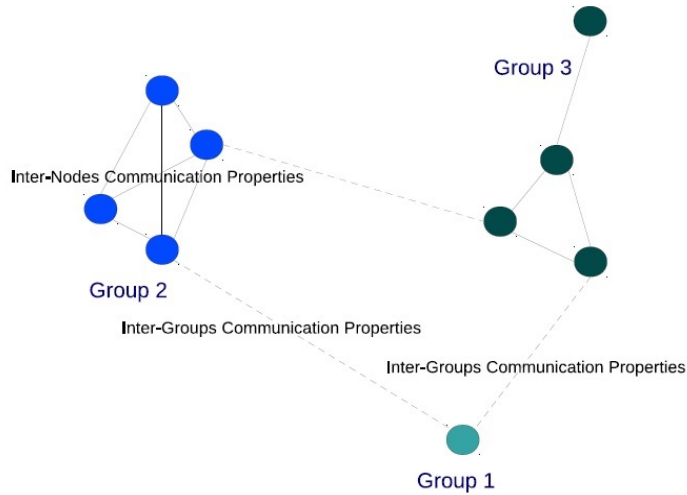

Figure 4: a) Extraction of the resource requirements from a given application graph, b) The resource graph of the qualified resources for a given query.

Algorithm 1 illustrates the RMC resource requesting mechanism when it is invoked by a PM. Upon receiving a PM request, RMC tries to conduct all the requirements of the given request using all the resources it has under control. When it fails to handle a request using the available resources in its own administrative domain (i.e., the collection of distributed resources, which are managed by a RMC instance) RMC starts to extend its global view of resources by invoking a RR component (see Figures 5 and 6). Consequently RR negotiates with other remote RP instances in order to get information about the required resources which are in the free-ownership-list (i.e., a list of free resources managed by an RMC instance, whose ownership is allowed to be transferred on request) of other RMC instances in the system. If those resources can be found in the free-ownership-list of other remote RMC domains, the local RMC would be able to add them to its own administrative domain.

```
Algorithm 1: Pseudo code of mechanism for resource requesting in RMC when PM asks RMC to
provide the resource descriptors for a set of required resources.
    Input: ProcId app, ResourceRequestList rl ;  /* process identifier and the description of the required
              resources */
    Output: ResIDs ;                              /* res-IDs for the matched qualified available resources */
    ResID-Collection reserved-resources; ResId i;
    ResID-Collection free-remote-resources; Resource res;
    ResourceRequestList rr = rl;
    ResourceRequestList request-remote-resources;
    ResourceType rt ;
    foreach rt ∈ rr do
        while rt.requested-number > 0 do
            i= find-free-resource(rt.descriptor) ;         /* searching in the local free-list and
            free-ownership-list */
            if i==null then
                ;  /* if the resources cant be found in the local set, the RMC will ask RR to find them
                in the system */
                 request-remote-resources.add(rt);
                break ;
            else
                dep=analyze-dep-constraints(i,app,rl);
                if dep is not achieved then
                    request-remote-resources.add(rt);
                    break;
                else
                    make-busy(app, i);
                    reserved-resources.add(i);
                    rt.requested-number- -;
    free-remote-resources=RR.FindResources(request-remote-resources);
    foreach res ∈ free-remote-resources do
        free-resources.add(res);
        make-busy(app, res);
        reserved-resources.add(res);
    return reserved-resources;
```

We must note that the resources, which are controlled by a single instance of RMC, are not necessarily positioned in a local node. In fact there, each RMC instance has three main resource collections, which include the reserved-list (or allocated resources), the free-list and the free-ownership-list. The reserved-list maintains the status of the resources allocated to processes. The free-list keeps the information of the resources available for upcoming processes. However, the other remote RMC instances do not have access to these resources. After releasing a resource, it will be moved from the reserved-list to the free-list. Generally, the resources in the free-list will be moved to the free-ownership-list after a certain period of time. Nevertheless, depending on the type of processes and their QoS requirements, the RMC can decide to keep some particular resources in the free-list for a longer time. This could be an important feature particularly for deadline-sensitive applications. Examples for this situation would be real-time applications (such as online gaming, voice over IP), where processes must guarantee response within specified timeframes (i.e., deadlines).

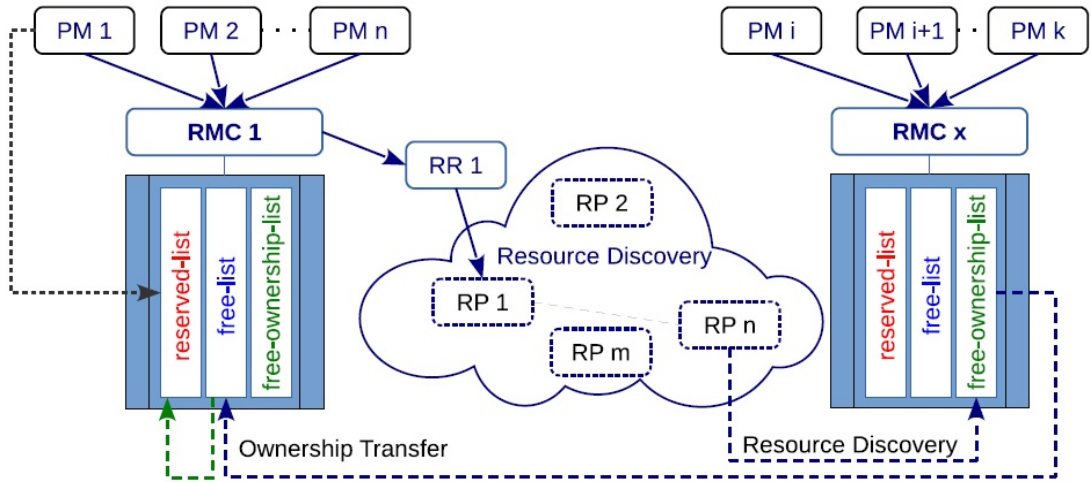

Figure 5: RMC mechanism.

This means that for such processes, correctness of an operation depends not only upon its logical correctness, but also upon the time in which it is performed [61] (i.e., a process may fail if has not met its specific deadline). RMC policy flexibility may allow longer maintenance of resources, released by deadline-sensitive processes, in the free-list, for two reasons: first, RMC potentially gets similar requests from PMs for upcoming processes, so that, RMC would be able to respond immediately to those requests without a need to perform a global resource discovery. Second, by keeping a resource in the free-list, instead of free-ownership-list, the scope for resource contention for the upcoming similar, deadline-sensitive, requests would be limited only to local RMC processes, and not to all processes around the system. This increases the probability of getting immediate response for those requesters.

The free-ownership-list contains the resources that the RMC will share with other remote RMC instances. Upon receiving a request for a resource, presented in the free-ownership-list, RMC transfers the authority to the requester (i.e., the remote RMC instance) to manage that resource. In fact, despite the free-list, the free-ownership-list contains the only available (free) resources which are allowed to be accessed globally in the system and their ownership can be transferred to the requesters on demand.

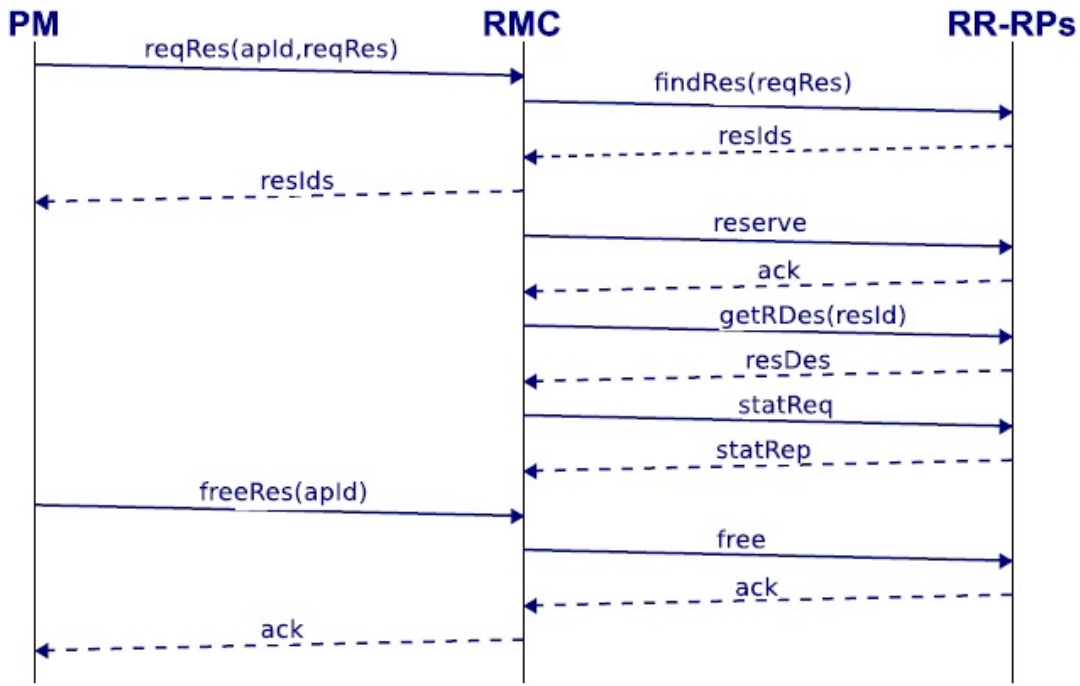

Figure 6: Message sequences for the interactions between RR, RP and RMC.

## 7. Resource Management System Operation

In this section, we elaborate on the inter-operations between the modules/services (which have been previously described in this report) in a general DOS environment and specifically for the discussion under the terminology for the project Service Oriented Operating System (S[o]OS) [58]. Figure 7 demonstrates general steps in configuration/inter-operation (i.e., in terms of modules, resources or threads), which contain all the major activities that must be performed for resource management and allocation (i.e., before the actual execution of the code's processes). These configuration/inter-operation steps include the following: a) Analysis of the program and extracting the application's graph. b) Identifying the resources required for a given process (i.e., identifying the resources, which maximize the matching between resource capabilities and process characteristics). c) Resource discovery, reservation, monitoring and management to find and reserve a matching graph of resources for a given application graph. d) Loading, configuring and instantiating all the necessary modules. e) Deploying the application segments in the reserved resources. f) Distributing code segments and their co-related reserved resources between the assigned sub-schedulers and starting to schedule the allocated threads on the reserved resources.

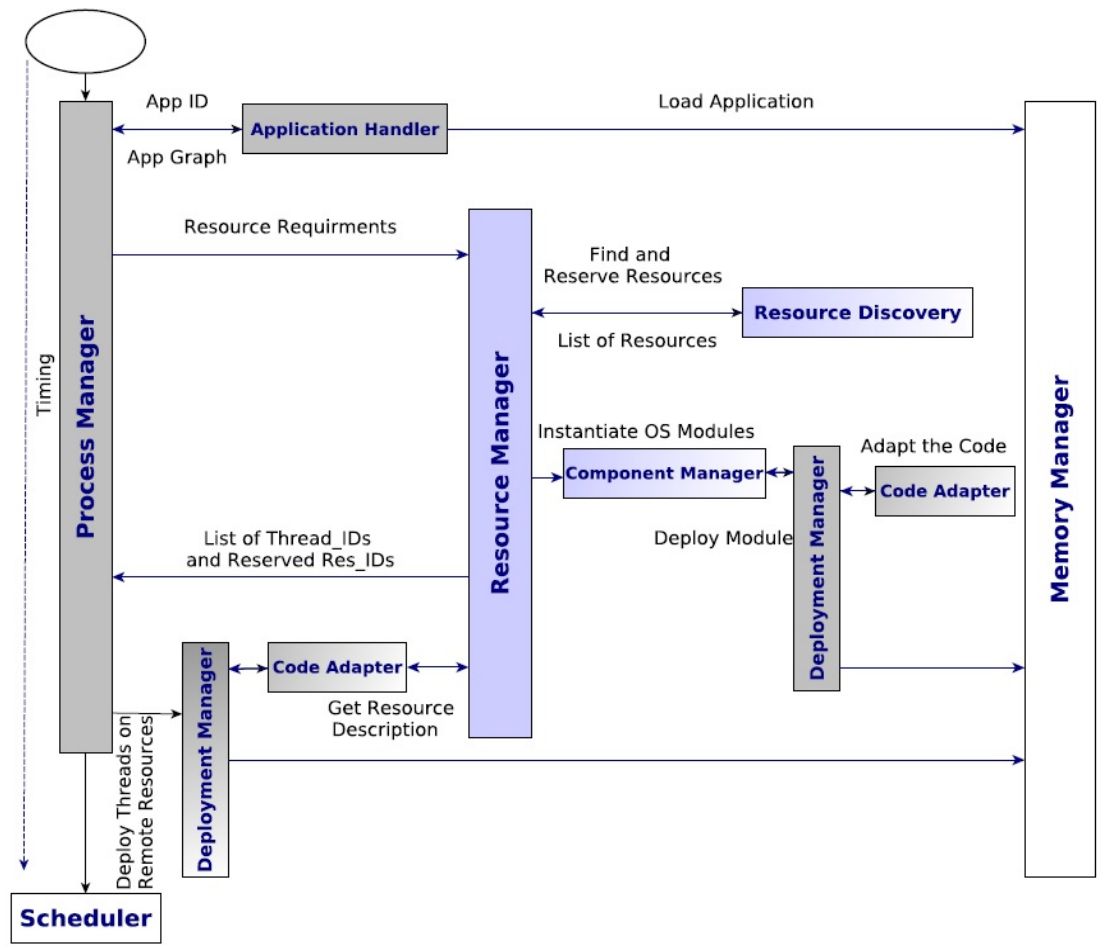

Figure 7: Interaction diagram among modules.

As depicted in Figure 7, the PM employs the application handler to load the requested application. The application handler invokes the memory manager (or stream manager) in order to load the application in memory and then it returns the application graph to the PM. In fact the code is analyzed with regard to its main dependencies, and the resulting application graph can be created with respect to the specific hardware features that would suit the execution best. Due to the application graph, the PM subsequently creates the list of resource requirements, which consists of all the individual characteristics of the potential resource candidate for each vertex combined with the aggregated characteristics and the required communication behavior of all the edges in the application graph. The PM will also pass the minimum and maximum resource requirements (i.e., the individual characteristics of the required resource in terms of processor architecture, type of processor, speed, etc.) for each thread, as well as the minimum and maximum communication properties among threads in terms of latency and bandwidth (i.e., the inter threads communication requirements) to the RMC. This means that RMC, regardless of the specific requirements of different applications, is able to discover and assign the required set of resources within given ranges (in terms of different computation or communication properties/attributes). In other words, RMC can be queried for resources satisfying arbitrary range conditions on different attributes since it benefits from a range-query enabled resource discovery mechanism. For example a PM may send a query to RMC with conditions such as the following: the required number of resources is 5, core clock rate for each resource must be greater than 1.5 Ghz, L1 cache size for each resource must be in the range of [128Kb, 1Mb] and also maximum communication latency between each pair of resources must be less than 10ms.

In current Grid and Cloud computing technologies, it is very common to specify the minimum and maximum computation requirements (i.e., desired computing attributes) for the resources requested by each query. Similarly, for many real applications, it is very important that the resource management and querying system can perform process-resource mapping with respect to minimum and maximum inter-resource/inter-process communication requirements. For example, for real-time applications, in order to meet the deadlines, we need to ensure that the maximum communication latency among the allocated resources does not exceed a given threshold.

The RMC specifies the resource graph and then it triggers the resource discovery module. The resource discovery starts to locate resources based on the required computing and communicating conditions. It analyzes all the possible resources and, in order to fulfill all the requirements of the input query, it chooses the most efficient and appropriate graph of resources, by considering the behavior of the communication routes between resources, which follows the communication dependencies (i.e., data source and sink) as identified in the application graph during code analysis. We must note that during the discovery procedure, every discovered resource (within the free-ownership-list of a remote RMC) will initially be marked as "reserved" for a thread in the application graph. This is the process which happens before transferring the ownership of the discovered resources to the local RMC (requester) and it results in a more immediate prevention of other remote instances of the discovery module to find the resources that are locally marked as "reserved" (i.e., the discovered resources for a query will immediately be hidden from other queries in the system and if not used in a given time window will be again released). It also must be taken into account that, in our approach, the discovery results (set of discovered resources for a request/process) are deterministic. This means that RMC does not need to either reevaluate the mapping conditions for the discovered resources or select the best matches among a set of candidates (discovery results), rather it simply assigns the discovered resources for the corresponding process.

Figure 8 shows an example of resource contention between two RMC instances which require the same resource and initiated at the same time. As we can see in the figure, the resource discovery requests from both RMC1 and RMC2 suddenly and simultaneously arrive at the same destination resource provider (RPn). RPn orders the incoming requests in its own FIFO queue based on the time of arrival (requests that arrive at the same time slot are organized in a generally random order). RPn processes requests one by one. Accordingly, the request from RMC2 will be successful in discovering its desired resource (resource Rx). The state of resource Rx will immediately be changed to "reserved", thus the next request in the queue (from RMC1) fails to discover the resource Rx. Afterwards, the ownership of Rx from RMCn will be transferred to RMC2.

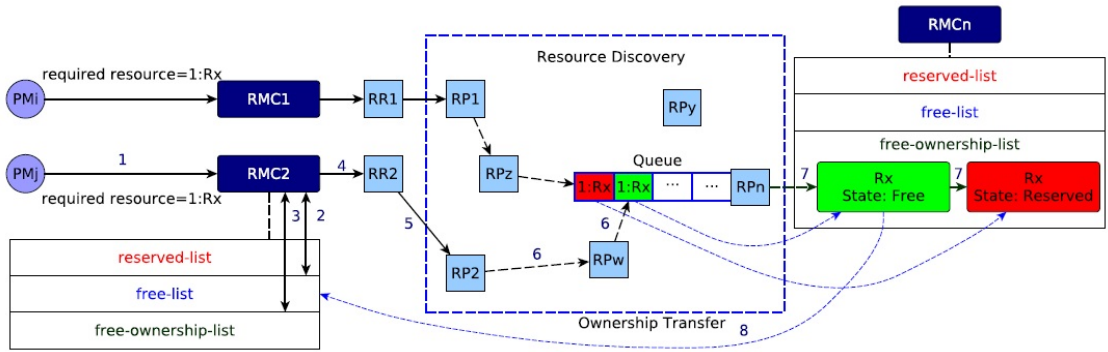

Figure 8: Example of resource contention occurs when two (or multiple) different RMC instances (like RMC1 and RMC2 on behalf of PMi and PMj) simultaneously attempt to obtain the same or overlapping set of rare resources (like Rx).

In fact, once the destination resources for application threads are known, they will be reserved and prepared. All necessary DOS modules for maintaining communication, loading code, executing rendering, etc., will be loaded along with the actual code. For doing this, the RMC will trigger the component manager in order to instantiate and setup all the necessary DOS modules. The component manager consequently employs the deployment manager to deploy the respective modules. The deployment manager has the resource description of the target hardware platform (remote resource) and it can invoke the code adapter to translate the code and finally write the respective adapted code to the target resource memory, by invoking the memory manager. Then the PM triggers the deployment manager in order to deploy each of the threads instructions in the allocated remote resources. In turn, the deployment manager calls the local RMC to get the full resource description for each resource, and afterwards it starts deploying each thread on the relevant remote resource platform. The PM will finally pass a set of threads on a set of reserved resources to the scheduler component in order to begin the actual execution.

## 8. Resource Management for FMCS

We can envision that, in the future, very large dimension manycore systems will be constructed by connecting a large number of computing nodes, with many chips and several thousands of heterogeneous cores per chip, using very high-speed network and interconnect technologies. In such FMCS, mapping applications to cores, and adapting each application to the allocated cores plays a key role for an efficient utilization of the computation resources. Efficient resource mapping and allocation in systems with thousands of cores integrated per chip spans a large solution space, and leads to the problem of optimal mapping of parallel applications to the cores which is known to be NP-complete [62].

## 8.1 RM Strategy

Our proposed resource management scheme can be leveraged for FMCS to provide a highly decentralized, distributed, scalable and online application mapping with highly resource allocation accuracy. The system is initially configured as follows:

A) In the first step, a virtual hierarchical overlay is created on top of the manycore system by dynamic self-organization of the vnodes (virtual nodes) over the system. Each vnode is defined as a group of homogeneous resources (i.e., processing cores) which are not necessarily positioned on a common physical chip (e.g., a CPU) rather they are positioned within a common vicinity that is described by parameters such as number of hops or interconnect latency (i.e., resources are grouped within virtual-nodes by proximity and similarity). We must note that the resources within a vnode can be also bound to a single common chip. Furthermore, the number of resources within a vnode is bound by a pre-configured threshold value. Therefore, the vnodes with similar resources (i.e., sibling-nodes) might coexist within a single common chip. All vnodes in the system are connected to each other through an overlay topology which is a connected graph created based on the underlying network and interconnect topology during the self-organization phase.

We define two main parameters to construct every vnode which are $\eta$ and $\lambda$. $\eta$ is the maximum number of resources (cores) in the vnode and $\lambda$ is the maximum distance between each pair of resources. The maximum distance can be also defined in terms of latency or number of hops. Depending on the values of the aforementioned parameters, the vnode clustering can be obtained through various approaches. It can be simply achieved by specifying each single core ($\eta=1$, $\lambda=0$), or all the cores of a homogeneous processor ($\eta$=number of cores in the processor, $\lambda$=maximum latency among cores in the processor, $\lambda < \delta$ where $\delta$ is minimum latency for inter-chip communication) as a vnode. Another way is to simply specify every group of homogeneous cores, positioned in different processors (i.e., chip or die) of a network node, as a single vnode ($\eta$=total number of cores in the node, $\lambda$=maximum latency among cores in the node, $\lambda < \Delta$ where $\Delta$ is the minimum latency for communication between cores of different nodes across the network). In fact a vnode is limited to include cores inside a single chip when $\lambda < \Delta$ and it is extended to multiple chips when $\lambda \geq \Delta$. Apart from that, various dynamic approaches can be also leveraged to perform vnode clustering. In the following, we briefly describe a sample approach for dynamic vnode clustering.

In this approach, for each multiprocessor node in the system, a random single resource (core) is triggered to initiate the clustering process. This resource, which is called primary vnode-head, would be responsible to organize the initial vnode for each physical node (multiprocessor node). The primary vnode-head broadcasts a vnode-clustering-request to all the cores in all the processors positioned in the current physical node and it will receive the description of each resource as well as the latency information from the cores. The primary vnode-head chooses a subset from the responding resources with respect to their similarity and vnode clustering parameters (i.e., $\eta$ and $\lambda$). In this way, a candidate resource of c is selected based upon three conditions: firstly the resource description of c must be matched to the primary vnode-head. Secondly the latency between the primary vnode-head and c must be equal or less than $\lambda\_$ and lastly by selecting c, the number of members for the primary vnode-head, should not exceed η. Upon establishing the first vnode, the primary vnode-head specifies the next vnode-head by randomly selecting a resource among the list of unsuccessful candidates. The next vnode-head will be triggered by receiving a message from the primary vnode-head, containing the list of unsuccessful candidates. The operation of the next vnode-head is similar to the primary vnode-head, but what makes it different is that instead of broadcasting to all cores, it uses multicasting to send its vnode-clustering-request only to the resources which were unsuccessful to join the previous vnode-head. The overall clustering process will continue until the list of unsuccessful candidates for the current vnode-head becomes empty. Figure 9 demonstrates an example of using this approach for vnode-clustering in a single chip containing 4 different types of resources. It results in establishing 5 vnodes.

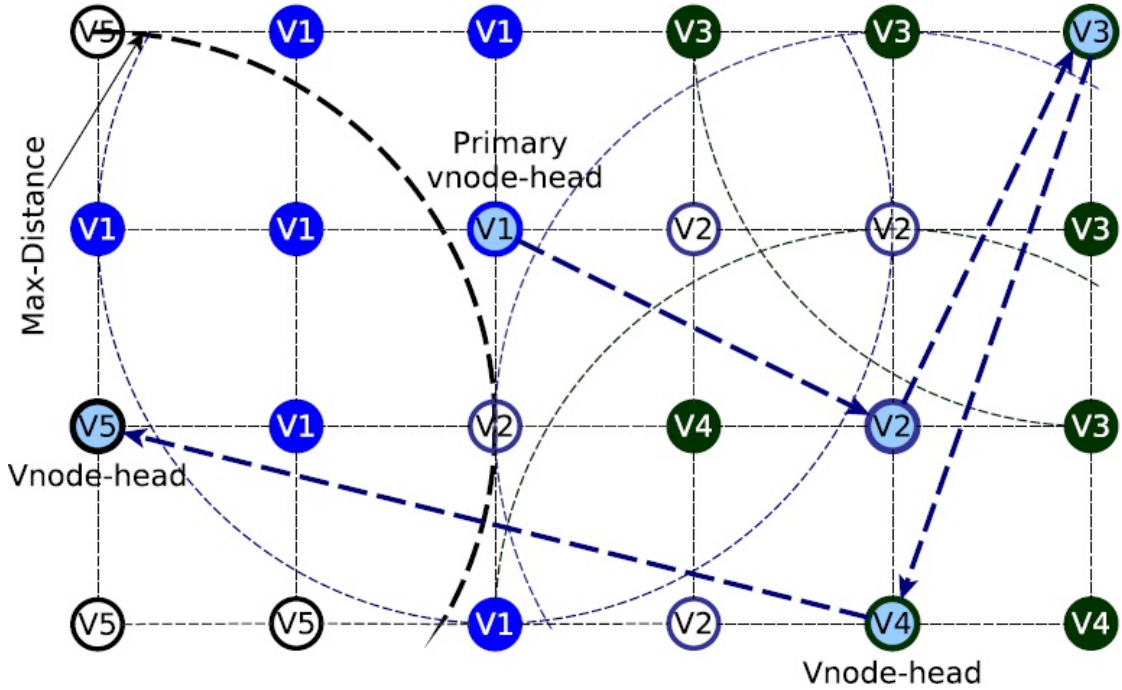

Figure 9: Example of vnode clustering for a processor containing 24 heterogeneous cores, connected through 2-D mesh topology

**B)** In the next step, each vnode acquires its role (LN:leaf-node, AN:aggregate-node, SN:super-node) to play in the computing environment using a dynamic distributed election mechanism. ANs and SNs are the resource providers (RPs) where ANs provide Query Management Service (QMS) and SNs provide Super Query Management Service (SQMS). These services work as the system services interleaved with the applications on the cores of each vnode with the AN or SN role. Alternatively, they can be also executed on the dedicated cores that are not available to the applications in each target vnode.

As we already discussed in Section 5, the vnodes which host QMS (i.e., RP-QMSs) are responsible to manage querying within a group of LN members and their-self. Similarly, the vnodes which host SQMS (i.e., RP-SQMSs) manage the querying within a group of AN members as well as their local resources. This mechanism divides the querying space to multiple layers such as layer-ln, layer-an and layer-sn where in each layer different adaptive query resolution and forwarding methods (such as DHT-based, Probability-based and Anycast-based approaches) are applied to direct and resolve the queries within or across layers. We must note that a single LN/AN is not allowed to participate simultaneously in more than one LN/AN group. Moreover, each compute node hosts atleast a single RP-SQMS, while every chip on a compute node hosts a single or multiple RP-QMS. Depending on the system configuration, the resources within a vnode or multiple vnodes initially managed by a single instance of RMC which cooperate with a RR entity. In other words, each RMC instance initially assigns a set of resources to its local free-list.

## 8.2 Application and Resource Description Aspects

We describe resources (i.e., computing resources) in the system based on their specific attributes (i.e., computation and communication characteristics) in a multi-layer hierarchy. The depth of the hierarchy (i.e., number of layers in resource description model) and the definition of each layer might range from very high level (e.g., super clusters, clusters) to very low level (e.g., processing core, ALUs) depending on the architecture designing aspects. The model is conducted by gathering and combining the individual attributes (ranging from more abstract information in the higher layers to more detailed characteristics in the lower layers) in each layer, augmented with information aggregation and summarization techniques, in order to create the layer-stamps. In fact, all specifications (i.e., attributes) of each layer, as well as their values are represented by a single layer-stamp. We also use a similar description model to specify the desired resources by each single query as <c|nln.c|nan.c|nsn>. This query-scheme identifies the status of the query conditions for each individual layer in the hierarchy. cln, can and csn denote the existence of query constraints for the layer-ln, layer-an and layer-sn accordingly, while nln, nan and nsn denote the non-existence of any query-conditions on those layers.

A detailed application description (i.e., application performance modeling) is necessary to accurately specify the desired resources, as well as the minimum/maximum required inter-resource communication capacities for the efficient application execution. By using an appropriate application description model, PMs would be able to dynamically query for the right set of cores for their corresponding processes. Since the application description is out of scope for this dissertation, we assume that the PMs are independently able to initially extract the number and characteristics of required resources, as well as the required inter-resource communication capacities for their corresponding processes at run-time and in an explicit manner. For example (see Figure 4), a PM for a process containing 9 threads with different resource and inter-thread communication requirements might ask the RMC to allocate 3 different groups of homogeneous cores to fulfil both the inter-group and intra-group constraints for communication between allocated cores. We also assume that applications/processes consist of tasks/threads that are executed on a single core and that may be freely re-allocated to different cores.

## 8.3 Matching Strategy

When a RMC receives a resource request from a PM, if the demanded resources are not in the local free-list and free-ownership-list, the RMC invokes a RR component to perform a resource discovery. RR in turn sends the given query to a RP-QMS, where the RR host (i.e. a vnode which hosts the RR service) is a member of its LN group. Due to the type of query and query demand (e.g., simple single resource, multiple heterogeneous resources, complex resource graph, containing constraints for inter resource communications) the RP-QMS splits the main-query in a set of sub-queries, and chooses individually the appropriate layer for each sub-query to start being processed. The query processing in layer-an is based on distributed probability tables that facilitate dynamic distributed learning processes, which are adapted to progressive environmental changes. In layer-ln, LN members participate in a specification-based DHT ring where the sibling nodes are linearly organized in linked lists, with single entries on the DHT ring, and where vnodes are positioned in DHT based on their core-level specification stamp. The query processing and forwarding in layer-sn also leverages the specification of the resources at the node-level, to conduct a specification based anycasting method and direct the queries among SNs.

The matching strategy of the resource discovery mechanism is based on matching the given query descriptions for each layer (desired attribute conditions for each layer) to their corresponding layer-stamps, at the time of visiting every individual vnode in different hierarchies. As we mentioned earlier, each hierarchy consists of three layers and every vnode depending on its role in the hierarchy (LN, AN or SN) provides layer-stamps. The exploration process for every sub-query (i.e., a sub-request, aiming to discover a set of homogeneous resources within a common vicinity) starts from the nearest, potentially qualified and available, resources to the far resources with respect to giving matching priority from the highest layer to the lowest layer (layer-sn to layer-ln). This way guaranties that the nearest resources (to the origin requester) are discovered in advance. Each sub-query may include a number of required resources with similar specifications as well as inter-resource communication requirements (e.g., maximum communication latency among desired resources). For each sub-query, in order to match the given communication requirements to the inter-resource communication links among the matched resources, we use the following approach:

1-Sub-query proceeds to explore vnodes/resources (from nearest resources to far resources), regardless of its communication conditions, focusing on computation requirements (query requirements for different layers).

2-Upon discovering the first match, sub-query sets it as the initial pivot (reference) for measuring inter-resource communications.

3- Sub-query continues to discover the rest of required resources and when it found the next match, it makes a decision, by first evaluating lower bound, and then evaluating upper bound of inter-resource communication requirements, to perform one of the actions listed: ignore the current matched resource, perform pivot switching or continue the normal search. Sub-query ignores the current matched resource and continues the discovery process, if the current matched resource fails to meet the lower bound requirement (minimum latency). If this doesn't happen, sub-query must next evaluate the upper bound condition (maximum latency). Accordingly, pivot switching might be required. Pivot switching is the operation for changing the pivot of a sub-query to the current matched resource (see Figure 10). By doing a pivot switch, sub-query ignores all previously matched resources and adds the current matched resource, as the first match, to its list of matched resources. Afterwards it proceeds to discover the rest of required resources. Pivot switching happens when the current matched resource fails to meet the upper bound requirement (maximum latency). If pivot switching doesn't happen, the discovery process proceeds normally, until all required resources are detected.

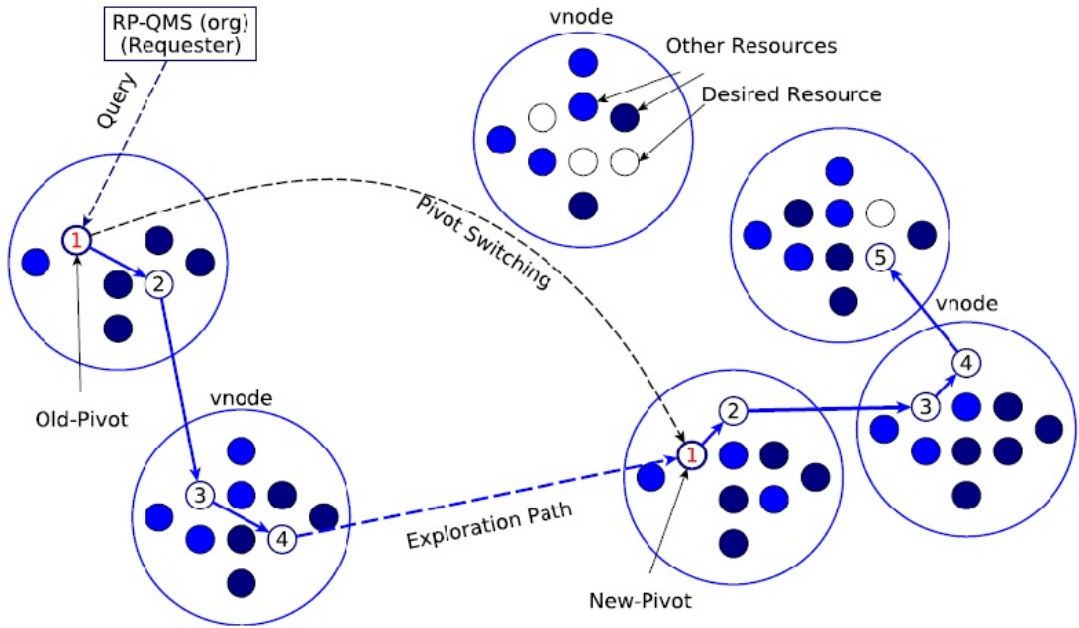

Figure 10: Example of matching strategy for inter-resource communication requirements using Pivot Switching mechanism. A sub-query, for 5 homogeneous resources with given maximum inter-resource latency, initiated from RP-QMS*org* and explores vnodes (from the closest vnodes to the most far one). The sub-query switches the pivot, when the recent matched resource fails to meet the upper bound communication requirement (maximum latency).

Finally, the RP-QMS that originated the sub-queries, aggregates the discovery results and replies to the RR with a set of resource matches that optimally satisfy the main-query demand. After acquiring the discovery result from the RR, RMC starts to perform a resource trading operation. It begins to negotiate with other related RMC instances in the system which maintain the discovered resources in their free-ownership-list and consequently transforms the ownership of those resources to its local free-list.

Each RMC instance periodically monitors the status of resources in its own resource pool, through a HotSpot-detector component which is responsible for detecting the overloading resources in the resource-pool. Whenever a HotSpot is detected, the RMC re-evaluates the resource requirements for the running application. Accordingly the RMC allocates set of new resources for that application, and starts to migrate processes to the new resources. This mechanism provides elasticity for ElCore to be able to tolerate the application with dynamic behavior (aka malleable applications).

## 9. Evaluation Results

In this section, we evaluate the performance of our proposed resource management scheme with respect to different evaluation criteria. We use simulation instead of experiment on the real large-scale computing infrastructures (e.g., PlanetLab, TACC, Oak Ridge, BSC, GENCI and public Cloud providers like Amazon and Google) due to low cost and flexibility of the simulation to design, development and evaluation of the new algorithms as well as providing full control over system behavior and evaluation scenarios. Furthermore, the real infrastructures generally provide limitations to explore the design space particularly for scalable performance and large-scale evaluation.

In general, one of the traditional evaluation criteria, is resource utilization, which demonstrates how the resources in terms of computation and communication capabilities (e.g., CPU usage, Memory usage, Network Bandwidth, etc,.) have been efficiently utilized, in order to distribute and manage the workload. Resource utilization as an evaluation criteria has been used for many resource management systems specifically for virtualization based resource management systems or Cloud platforms available today. In our evaluation, we decided to use the Resource Allocation Accuracy (RAA) as the main criteria to evaluate our work instead of Resource Utilization. RAA indicates how much a set of allocated resources for a process fulfils the original resource requirements, emphasized by the PM requester. In our resource management scheme, we assume that PM would dynamically be able to precisely identify the resource requirements for each process and consequently the next level RMC would be responsible for finding and reserving the optimum set of qualified resources, which will be allocated to the relevant threads by PM, where the scheduler will be invoked to run the process accordingly.

We note that our assumption of dynamic identification of resource requirements for given applications/workloads (by PMs), may not be fully practical in current manycore systems (or implemented in current operating systems), but it is becoming a practice in nowadays Cloud computing technologies [63–74]. We argue that current Cloud computing systems are a potential ancestor of the future manycore systems (Intel 48-Core "Single-Chip Cloud Computer" [75] is an example of this trend). Since the target computing environments of ElCore is the future manycore systems, our assumption is reasonable; this is further supported by the concepts developed in the context of S[o]OS (which is an example of distributed operation systems, concerning the requirements of future manycore systems) [76]. In other words, S[o]OS is not for today, but we can see the ancestors of S[o]OS on current cloud systems, where the service demand (or the resources needed to run a job) is estimated prior to VM allocation and job execution [63–67].

Due to the above assumption, and the fact that the efficient mapping between the RMC offered resources and the PM resource requests has significant impact on the whole system performance, we consider RAA as one of the most important performance criteria for managing resources in future large-scale manycore environments (with presence of high diversity of resources and applications) since it fully depends on the quality of resource management component, while resource utilization very much depends on the specific dynamicity of the requests being processed as well. As it is shown in Equation 1, RAA for each query can be measured by calculating the ratio of the fully satisfied conditions to the total number of query conditions (in terms of the required computing attributes for each single resource, and the desired inter-resource communication properties).

$$RAA = \frac{\gamma + \sum_i^g \left( \sum_j^{m_i} (\phi_j) + \tau_i \right)}{L + \sum_i^g (n_i \rho_i + l_i)} \times 100 \qquad (1)$$

Here, $\rho_i$ is the number of desired (computing) attributes for each requested resource in a sub-query for a group of homogeneous resources (i is the group index); $n_i$ and $m_i$ are respectively the number of requested resources and discovered resources for each group i; $l_i$ and $\tau_i$ are respectively the number of dependency links and qualified links for each group; L and $\gamma$ are respectively the number of inter-group links (communication conditions) and qualified links for all groups; g is the number of resource-groups (number of sub-queries for the query) and $\varphi_j$ is the number of qualified attributes for a discovered resource (j is the resource-index).

9.1 Simulation Setup

To do our evaluation, we developed a simulation platform, based on OMNET++, which is able to simulate manycore environments (up to 55000 cores in different chips and nodes), focusing on communication aspects (i.e., communications between cores, chips and nodes). Using our simulator, we have simulated a manycore networked environment containing 2000 computing resources (i.e, processing cores), with 6 different types of processing resources. Each of these types have their own specific computation properties in terms of core clock rate, cache size, cach line, ALU properties and functionalities, memory bandwidth, etc. For instance, the processor frequency for resource type A, B, C, D, E and F are 2.53 GHz, 3.6 GHz, 1.6 GHz, 2.8 GHz, 1.2 GHz and 2.4 GHz accordingly. We conduct our simulation in a way to produce 20 percent of the resources as distributed homogeneous resources of a specific type-A which later will be required for several different processing scenarios. The reason for adjusting the amount of a certain resource-type in the system (as the target resources for the queries) is to facilitate measuring the value of Maximum Reachable Accuracy (MRA) in the different experiments, but in fact it does not impact the generality of our evaluation because the resource management procedure is completely independent from the frequency of the target resources.

The type of resources, as well as their distribution in the system, is random. To do this, in the first step, we randomly order all processors (of different nodes) in a queue. We create an array of pairs (tx,ty) which contains all different two-combinations from the given set of types S={A,B, .., F}. We iterate through the array, and for each iteration, we dequeue a processor from the queue and proceed to randomly specify the type of tx or ty for each of its cores. The iteration is repeated and the process continues until the queue becomes empty. As a result, each processor in the system, will consist of maximum two different types of cores (heterogeneous cores). In the next step, we control and regulate the desired amount of cores of a given specific type (type-A) through modifying the type of cores from type-A to ty or from ty to type-A, in the processors that contain cores with type-A (tx=type-A).

Figure 11 shows the overall architecture of the simulated manycore system that we use for the evaluation in this section. The cores of each processor are connected through a three-dimensional mesh interconnect topology (with datarate of 50Gbps) which, in our simulation scenario, is the same as three-dimensional torus. Each core has its own dedicated L1 and L2 cache that are not shared with any other cores. We do not simulate cache coherency protocols, as we use message-passing for inter-core and inter-chip communication. The routing among cores is performed through a variation of Dimension Order Routing algorithm [77] where packets are routed to the correct position along higher dimensions (x or y) before being routed along lower dimensions (y or z). Furthermore, the processors of each node are connected through a high speed bus with datarate of 20 Gbps and we construct a network (with bandwidth of 100Mbps) based on a random network topology, a connected graph with $\beta = 62.5\%$ that indicates the ratio of the number of nodes to the number of links in the network graph.

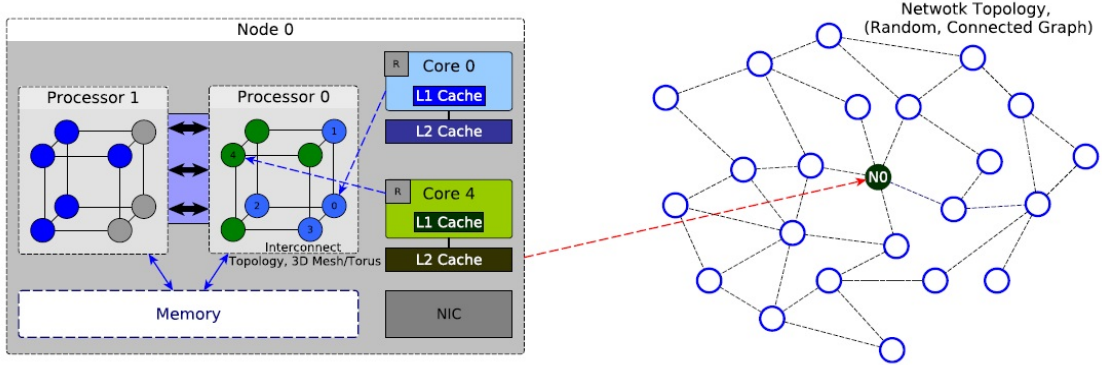

Figure 11: Manycore Simulation - System Architecture

We define 4 settings for the processing scenarios. Every setting is evaluated in 10 different runs. In each run the network topology parameters will be randomly changed according to our predefined simulation parameters (see Table 1).

Table 1: Simulation Parameters for Accuracy Evaluation

| Parameter | Values | Description |
|---|---|---|
| Physical network size | 2000 cores | Total number of resources in the system (i.e., total number of cores) |
| Number of network nodes | 125 nodes | Total number of connected network nodes in the system |
| Number of tiles (CPUs) per node | 2 processors | Number of processors per network node |
| Number of cores per CPU | 8 cores | Each processor contains maximum two different types of cores |
| Interconnect topology | 3-Dimensional Mesh/Torus | Topology among cores of each processor |
| Network topology | Random (connected graph, $\beta$=62.5%) | Topology among network nodes, $\beta$ indicates ratio of #nodes to #links in the network graph |
| Interconnect channel datarate | 50Gbps | Inter-core communication |
| Routing type | DOR | Dimension Order Routing algorithm for routing among cores |
| Bus speed | 20 Gbps | Inter-processor communication |
| Network channel | 100 Mbps | Inter-node communication |
| Vnode clustering | $\eta$=8, $\lambda$=100ns | Clustering is limited to include cores inside multiple chips of a single common network node |
| Desired inter-resource latency | (0,150$\mu$s], maximum 150 $\mu$s | The maximum inter-resource latency required for each query |
| Process Duration by sec | Weibull($\lambda$=3.58,$k$=2.40) | Execution time for each process through Weibull distribution, $\lambda$ is the scale parameter and $k$ is the shape parameter. |

In the first setting we assume that PMs will generate 1000 resource requests in fixed intervals (4000 ms) for different processes of the same type (i.e., in each interval, only a single request is generated in the entire system). The predefined process in this setting includes 2 threads that need 2 resources of type-A with specific communication requirements between those resources (maximum 150 μs for each dependency link). The other settings are similar to the first setting except for the type of processes. However, for the sake of simplicity, the type of requested resources in all of these settings are homogeneous (type-A). The difference between the processes in the aforementioned settings is the number of requested resources, as well as the level of dependency between the threads. We also assume that the required resource for each thread is a single processing resource (i.e., a core). The required number of resources for each process in settings 2, 3 and 4 are 3, 5 and 9 while the threads dependency constraints are applied for 2, 4 and 8 communication links accordingly (see Figure 12).

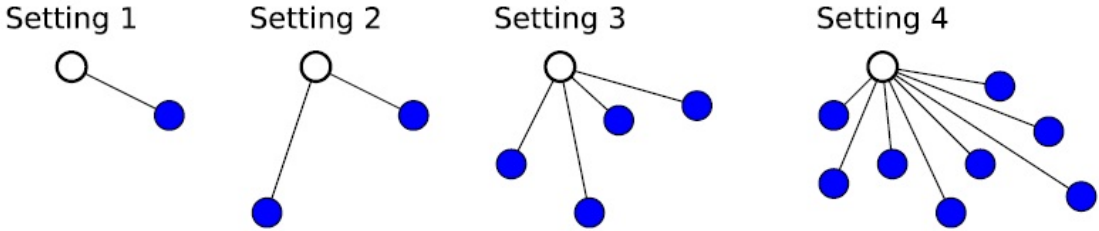

Figure 12: Dependency graph for different settings

In each run, after a query is finished, the matched resources will be allocated to the corresponding process. The allocated resources will be released after a period of execution time (termination of the process, see Table 1). In our simulation we assume that the released resources will be immediately transferred from the free-list to the free-ownership-list. For each resolved query, the simulator records information including query description, query results and the current status of the network (system) graph as a trace. Upon completion of simulation, the simulator individually analyses the generated traces for each run in order to calculate values of RAA and Maximum Reachable Accuracy (MRA) for each query. MRA is measured, similar to RAA (as it is presented in Equation 1). But, unlike RAA, MRA uses the parameters of the ideal query results, instead of the real query results achieved by our proposed resource management approach. To do this, the simulator statically evaluates the query conditions for each potential set of results in the network graph (provided by a trace) and concludes with the most optimal results attainable for the given query.

## 9.2 Resource Allocation Accuracy (Mapping Quality)

Figure 13 demonstrates the simulation results for each of the defined settings for different runs. The linear plots above the bar charts in each graph represent the variation of the value for MRA which is dependent on network topology and random distribution of resources.

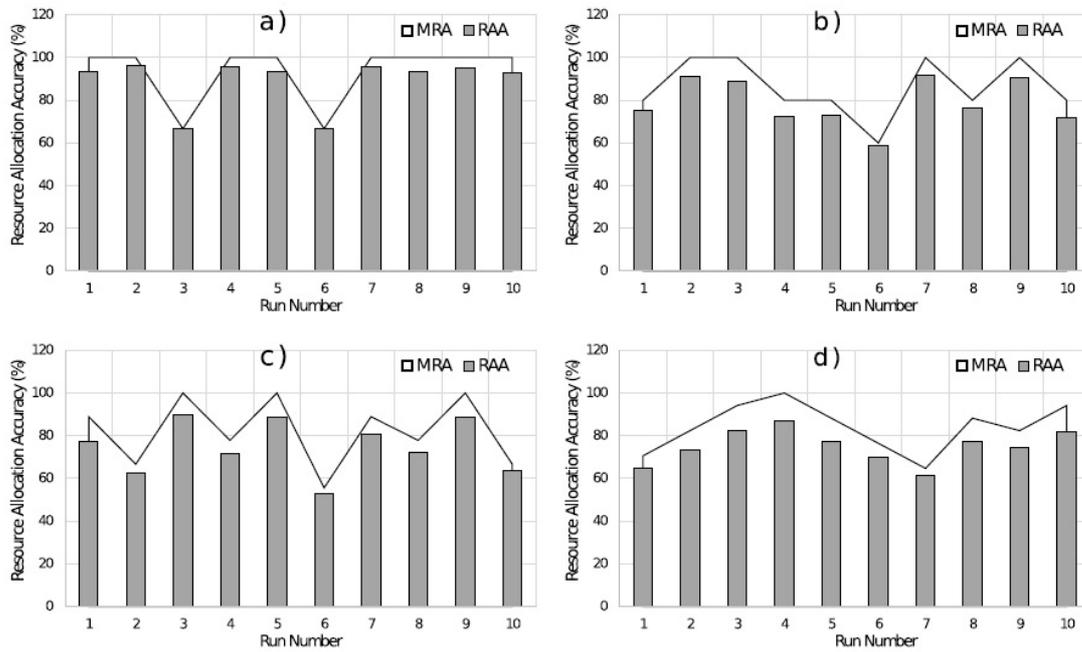

Figure 13: RAA ratio for different settings and network topology parameters: a) Setting 1: the processes with 2 threads b) Setting 2: the processes with 3 threads c) Setting 3: the processes with 5 threads d) Setting 4: the processes with 9 threads (MRA is the Maximum Reachable Accuracy, each run indicates different random network topology parameters).

As we mentioned earlier, MRA can be calculated by analyzing the states of resources and the network topology in each run. In fact, it is possible that request requirements for a set of resources not be fully obtained in the system. This might happen when the requested set is not available (reserved), or never existed in the system. In such a case our approach tries to offer a similar set, which can meet most of the requirements. MRA specifies the border line to offer the most optimal set of resources for a specific process due to the realistic conditions of the resources in the system. The bar charts represent the average percentage of RAA with different settings and runs. As can be seen in the results, the RAA in all of the tests are almost close to the value of the maximum reachable accuracy which presents a resalable level of accuracy for the resource allocation in the proposed resource management approach. The result shows that, the RAA for small processes (i.e., the process with small number of threads) (Figure 13.a and Figure 13.b) in comparison to the larger processes (Figure 13.c and Figure 13.d) is closer to the maximum in overall.

Figure 14 shows the average inaccuracy ratio for different types of processes with various levels of dependency between their threads. It can be seen that the inaccuracy ratio grows for the larger processes with higher level of thread dependency. However the inaccuracy slope is still less than the dependency slope. The reason for this behavior is that larger processes, with higher dependencies, have more requirements for their required set of resources, which can't be potentially fulfilled in the system in comparison to the smaller processes with lower dependencies. When the system is not able to provide the qualified set of resources considering all the process requirements, it returns the optimal results that partially meet the requirements. And this increases the inaccuracy ratio. Our approach tries to provide the most qualified set of resources for allocation. However if it is not feasible to obtain, (for reasons such as unavailability of the resources and system limitations) it offers the best possible set of resources.

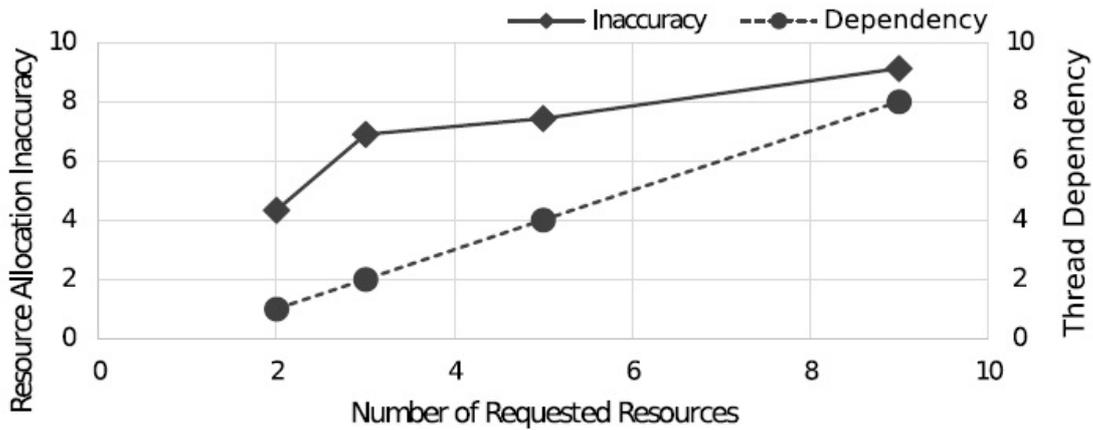

Figure 14: Resource Allocation Inaccuracy ratio for the processes with different number of threads and threads' dependency.

## 9.3 Scalability and Performance

In the rest of this section, we evaluate the performance of our resource management model with respect to scalability. To do this we conduct a simulation scenario based on the simulation parameters mentioned in Table 1 with the following changes, presented in Table 2.

Table 2: Simulation Parameters for Scalability Evaluation

| Parameter | Values |
|---|---|
| Requested Resources for each Process | 20 |
| Frequency of Target Resources (FTR) | 1650 |
| Physical network size | 27500 cores |
| Process Duration by sec | Weibull($\lambda$=3.58, $k$=2.40) |
| Querying Interval by ms | Exponential($\beta$=4000) |
| Consecutive Query Runs per RMC | 1000 |
| Number of Active RMCs | 275 |

We simulate a distributed dynamic computing environment containing 27500 computing resources, in which a constant number of RMCs (i.e., active RMCs) simultaneously handle the incoming resource requests from their PM clients. The time interval between each pair of consecutive queries issued by PMs (i.e., the arrival rate of incoming resource requests from PMs) is defined by Exponential distribution. We also assume that each active RMC only receives 1000 consecutive resource requests from PMs over the simulation time. RMCs assign and reserve the proper resources for each PM request. The resources reserved for each process be released after the execution time, which is defined by a Weibull distribution. We measure the system overhead caused by resource management components (i.e., number of transaction messages between RMCs, RRs and RPS to handle a resource request issued by a PM) as well as the RMC latency for requests over time.

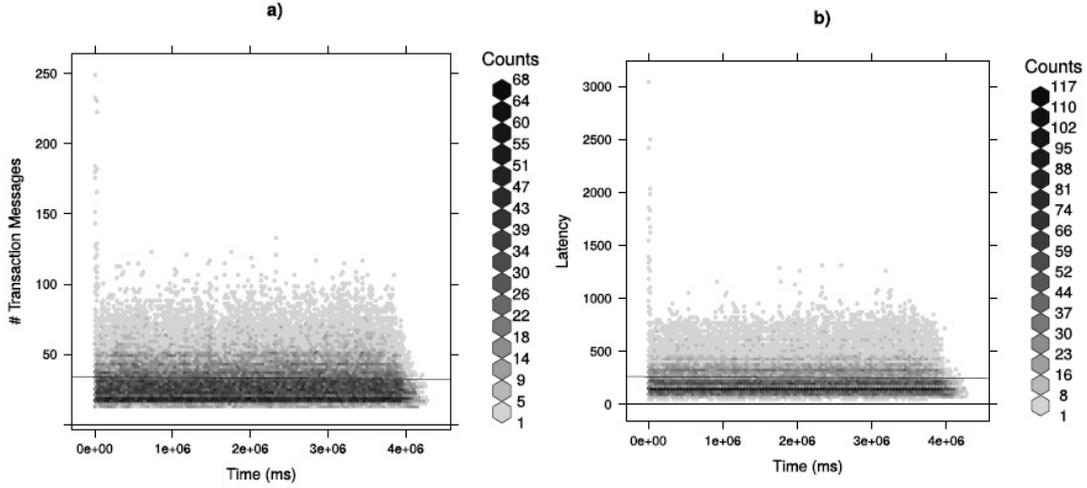

Figure 15: Hexbin plots for the number of transaction messages and the latency for each resource request over time.

Figures 15-a and 15-b present the Hexbin plots for resource management overhead and latency over simulation time (by millisecond). Each data point in the graphs represents the result of a single PM resource request. The darker points in the graphs (i.e., the high density points) represent states that have a higher probability of occurrence in comparison to the lighter points. As we can see in these graphs, the majority of the queries results, particularly the high density data points, fall on or below the regression line. This illustrates that our resource management scheme is highly scalable over time and can efficiently maintain its performance under natural

churn caused by high frequent resource reservations and resource releases in highly dynamic computing environment.

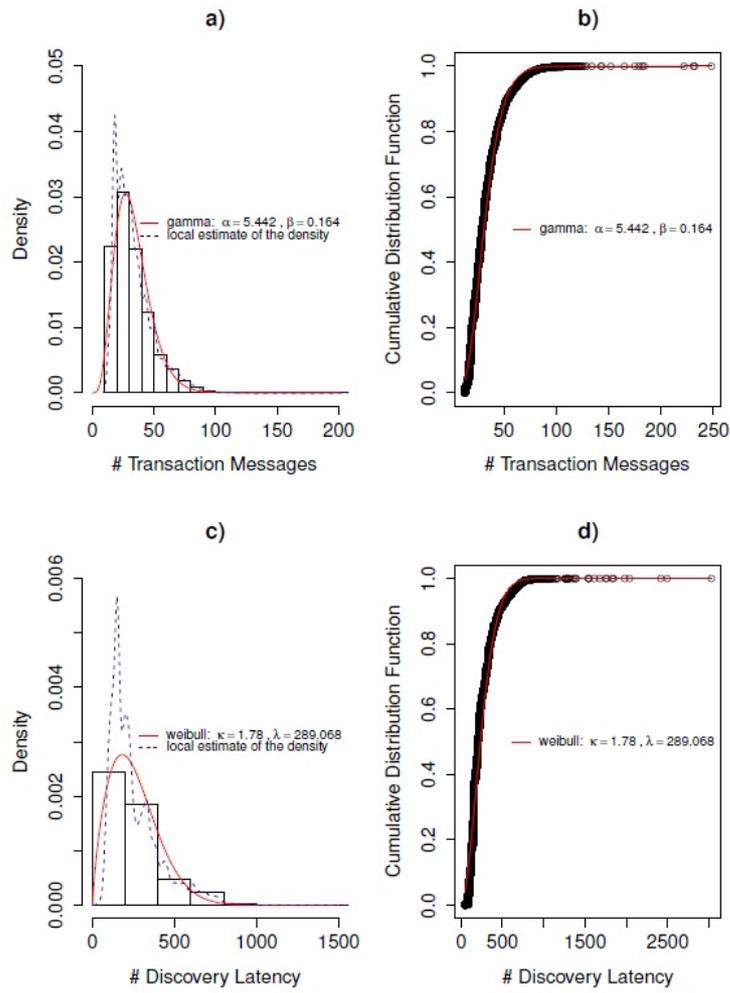

Figure 16: a and b) Histogram, density plot and CDF plot for the number of transaction messages between resource management components over simulation time, c and d) Histogram, density plot and CDF plot for the querying latency for the PM requesters over time.

Figure 16-a and Figure 16-b illustrate that at least 80% of the resource requests are managed by their correspondent RMCs with less than 50 transaction messages which shows a reasonable overhead. But for less than 20% of the resource requests, the resource management overhead is a value between 50 to 250 messages. This happens mainly due to resource unavailability, rare resources and resource contentions. In fact, for larger process execution times, it takes longer for the reserved resources to be released by the original requesters and this leads to higher rate of unavailability for the occupied resources in the system. In this regard, the frequent resources in the system, might become rare resources over time with higher cost of discovery. In addition, the rate of resource unavailability and the speed with which the resources become rare will accelerated particularly when the overall task duration (i.e., process execution time) exceeds the overall querying interval. Figure 16-c and Figure 16-d also present the similar results for the average querying latency to process the requests that were initiated by different PM components in the system over simulation time.

### 9.4 Scalability and Resource Discovery

In continuation of this section, we present the simulation results which demonstrate the performance of our resource management scheme with respect to scalability in comparison to other alternative approaches. The scalability of our resource management model is critically dependent on the deployment of a proper resource discovery mechanism that can discover resources managed by different independent distributed RMC components around the network. For comparisons, we simulate our model, ElCore, in conjunction with three other approaches which benefit from different generic hybrid distributed methods for resource discovery. These approaches are PRW, FRW and Broad-Walk. We also simulated another version of ElCore which employs a non-anycast based method for resource discovery. This model is identical to ElCore except that it doesn't support SN layer and SQMS providers for resource discovery and instead it resolves any SN-dependent queries by extending the native probability mechanism of ElCore to support the layer-sn resource information within layer-an.

In order to perform resource discovery, PRW (or partial random-walk), FRW (or full random-walk) and Broad-Walk are organized on top of two-layered (i.e., leaf-node layer and aggregate-node layer) distributed hierarchies. PRW and FRW leverage the same Chord based DHT method which is used in ElCore in the leaf-node layer, however, they provide different query-forwarding methods in the aggregate-node layer. PRW uses both probability and random-walk methods to guide queries in layer-an. In this approach, distributed probability tables in the system only process the query results with respect to the resource information in layer-ln and layer-an. PRW might behave similar to ElCore-nAC for non-SN-dependent queries (e.g., <cln:can:nsn>) but for SN-dependent queries the selection of the forwarding destination node is partially random since the probability tables do not actually care about the required resource conditions in the super-node layer for the given queries.

The PRW is comparable to our approach (ElCore) in the sense that it creates clusters on top of the unstructured network. It also provides similarity with some well-known request propagation strategies in the literature such as the shortcut, random walk, learning-based, best-neighbor, learning-based + best-neighbor methods. These methods have been used in many popular resource discovery systems and applications [78–83]. For example in Iamnitchi et al [84] a fully decentralized discovery approach is proposed which is based on publish/subscription of the resource information on some specific nodes in the virtual organization. Learning-based and also random-walk methods are used to propagate the queries among the server nodes. Our approach and also PRW are not based on publish/subscription since it is costly in terms of network traffic, processing, and storage needs for periodical updating and the maintenance of resource information particularly in high dynamic environments. On the other hand, our probability mechanism is comparative to or even better than learning-based strategy. In the learning-based method nodes learn from experience by recording the requests answered by other nodes (i.e., by caching the results of successful queries). A request is forwarded to the node that has answered similar query previously [85]. This strategy becomes inefficient when the system size, dynamicity and heterogeneity of resources/queries increase due to the larger memory requirements to maintain the query results and unavailability of the pre-discovered resources. But in our proposed probability mechanism, the statistical information about all the transacted queries by each peer are aggregated in the fixed-size DPTs regardless of the successfulness of the queries. In addition, by leveraging techniques such as dynamic best SOR detection, low-resource nodes and resource unavailability detection and various situation-based policies and updating strategies (e.g., shortest path, latency aware and attribute-based updating) our proposed probability method provides better accuracy and efficiency.

Unlike PRW, FRW employs a fully single random-walk method to guide all types of queries in aggregate-node layer. Random-walk is a common query-forwarding method which is originally proposed in the literature to alleviate the excessive traffic problem caused by flooding [86] and to deal with the traffic/coverage trade-off. Random-walk is used in many distributed resource discovery applications such as Gnutella [87, 88], Iamnitchi et al [84] and [89–93]. Broad-Walk is a hybrid two layered approach which uses broadcast-based query propagation method [94, 95] in the leaf-node layer and the random-walk forwarding in the aggregate-node layer. In continuation of this section we explain the details of our simulation setup and we discuss the comparison results.

Using our self-organized clustering algorithm we simulate the aforementioned discovery approaches on top of either two-layered or three-layered distributed hierarchies. Similar to the previous scenarios, we simulate dynamic computing environments containing various numbers of computing resources in which a constant number of resources (i.e., requesters) simultaneously issue the discovery requests to the system. The time interval between each pair of consecutive queries issued by a requester is defined by an exponential distribution. We also assume that each requester issues 10 consecutive resource requests to the system over the simulation time. The discovered resources will be reserved for each discovery request. The reserved resources for each process will be released after execution time period which is defined by a Weibull distribution. We execute 10 queries per requester for each system size, originating from a uniformly chosen source-node. Each experiment for each system-size is repeated for 100 run numbers with different topology parameters. All the queries are identical and represent the queries of type <cln:can:csn>. Each requester is willing to find required resources for a process containing three thread-groups with different resource requirements. Table 3 presents more details of simulation parameters for our evaluation.

Table 3: Simulation Parameters for Performance Compression

| Parameter | Values |
|---|---|
| Physical network size | 5500-55000 cores |
| Interconnect topology | Mesh/Torus |
| Network topology | Random |
| Interconnect channel datarate | 50Gbps |
| Network channel | 100 Mbps |
| Desired Resources for each Request | 3×20 |
| Homogeneity rate of desired resources | 33% |
| Frequency of Target Resources (FTR) | 1650 |
| Process Duration by sec | Weibull($\lambda$=3.58,$k$=2.40) |
| Querying Interval by ms | Exponential($\beta$=4000) |
| Consecutive Query Runs per Requesters | 10 |
| Rate of Requesters | 1% |

In the first test, we perform experiments to measure average number of required messages and average latency (by milliseconds) per discovery request for ElCore, ElCore-nAC and other approaches. The PRW, FRW and Broad-Walk with the same topology, simulation parameters and conditions are used as alternative reference works. Figs 17-a, b and c show plots of the average discovery messages and the average discovery latency per query as a function of the number of computing resources in the system (i.e., system size). Figure 17-b demonstrates the results presented in the Figure 17-a with better resolution (without Broad-Walk).

In Figure 16-a, we observe that the average required number of discovery messages per query for Broad-Walk is much larger than the other approaches while in Figure 16-c the average latency of Broad-Walk is almost close to FRW, PRW and ElCorenAC. This means that Broad-Walk significantly generates more discovery traffic in comparison to others due to the heavy cost of broadcasting in layer-ln. The queries in Broad-Walk are guided in the aggregate-node layer by being forwarded to a non-visited single random neighboring aggregate-node. Upon arrival of a query in an aggregate-node if that node fits the query conditions (i.e., can) for the current layer (i.e., layer-an) the query is broadcasted to all the leaf-node members of the current aggregate-node otherwise it is forwarded further in the network using random-walk. Broadcasting results in increased traffic but it could also provide reasonable response time for queries (as seen in Figure 17-c) since the aggregate-node inquires all of its leaf-node members in parallel. The response time for Broad-Walk is approximately close to the results for FRW, PRW and even ElCore-nAC.

Figs 17-b and 17-c show that our approach, ElCore, provides the highest performance and scalability among others for both discovery traffic (i.e., average number of discovery messages propagated during a search) and latency when varying the number of resources in the system from 5500 to 55000 resources. This is particularly significant for the query response time (latency) since the other approaches such as ElCore-nAC and PRW also provide close results in terms of the number of discovery messages. ElCore efficiently divides the exploring space to anycast groups in a way that queries with csn requirements are only propagated among the SQMS providers whose specifications in layer-sn fulfill the csn conditions of the given query essentially. In comparison to the ElCore-nAC, this strategy leads to a significant reduction in the response time of ElCore while its discovery traffic slightly decreases. As we already discussed, ElCore is the enhancement of ElCore-nAC by leveraging our proposed anycast-forwarding mechanism in an extra layer which is called layer-sn.

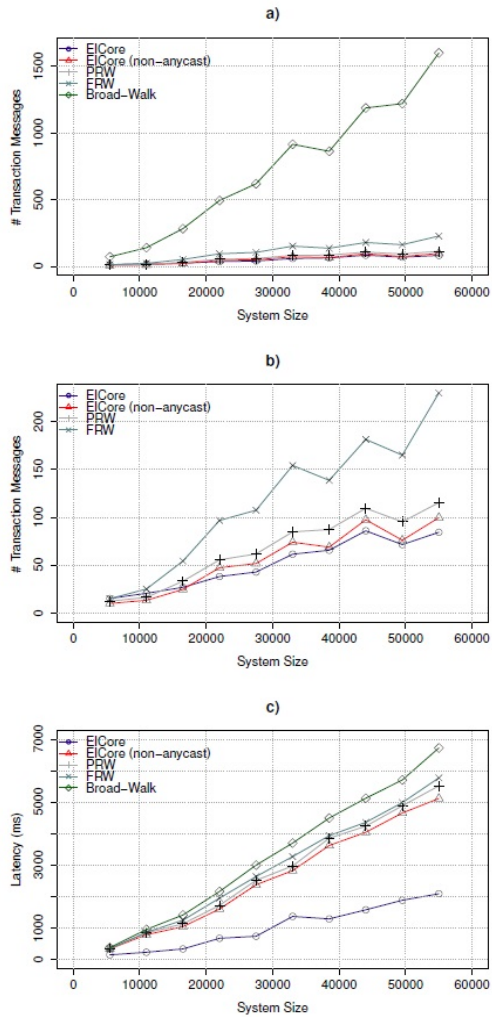

Figure 17: Comparison between ElCore and other alternative approaches: Average number of transacted messages and latency per query for different system sizes and different approaches.

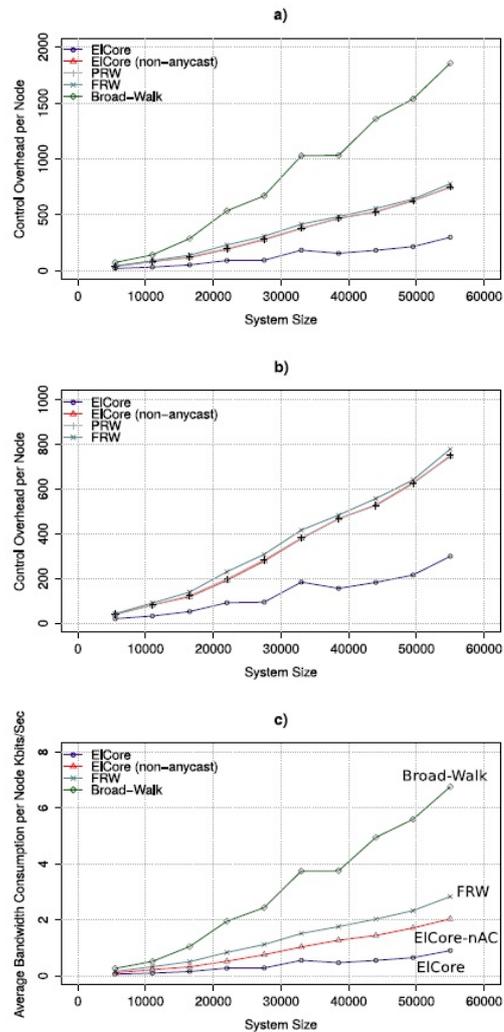

Figure 18: Comparison between ElCore and other alternative approaches: a,b) Average control overhead per node (i.e., average number of transmitted messages per vnode) and c) Average bandwidth consumption per node, during simulation time (60000-80000 ms), for various system sizes.

The presented results for ElCore-nAC and ElCore in Figs 17-b and 16-c also prove that increasing the level of hierarchy along with the implementation of an efficient adaptive corresponding query-processing method improves the overall performance of our discovery system. Another factor contributing to ElCore's overall performance is that ElCore controls the discovery procedure in a more intelligent way which saves much unnecessary message cost. Moreover due to the anycast nature of ElCore, SQMS providers are able to effectively guide the given queries to the closest qualified SQMS provider in the system which results in a significant reduction in the discovery latency for the queries in the system (see Figure 17-c).

From Figure 17-b, we can also see that, PRW generates larger number of discovery messages per query than ElCore-nAC and ElCore because of its partial random-walk query-forwarding mechanism in layer-an. In fact, PRW provides an efficient probability mechanism (similar to ElCore-nAC) to guide queries in layer-an to the potential matched resources in the system. But this probability mechanism becomes inefficient for processing the queries with csn requirements because the DPTs in PRW do not consider the csn requirements of the given queries in order to statistically estimate the target aggregate-node for query forwarding. This leads to a sort of partial random-walk for the SN-dependent queries (i.e., for the queries that csn <> nsn). But for the other types of queries (e.g., < cln:can:nsn >) which are not considered in our evaluation in this section, PRW is expected to behave identically to ElCore-nAC. Figure 17-b illustrates that FRW provides a lower performance with respect to discovery overhead compared to PRW while they exhibit almost similar behavior for discovery latency as shown in Figure 17-c. This is due to the fact that the mechanism for query resolution in layer-an of FRW is fully based on random-walk method. This means that the queries in layer-an are forwarded to a uniformly random selected neighboring aggregate-node in the system which is not yet visited. Since the next-node selection strategy is completely random-based the number of required traversed discovery messages for resolving a query gets more compared to the approaches benefiting a type of estimation-based strategy.

In the next experiment, we measure the overall discovery load per node (i.e., vnode) and the average bandwidth consumption per node during the querying period (60000-80000 ms), which is the duration of time that the requesters propagate a constant number of successive queries across the network in a parallel manner. The querying period ends when the request-initiator corresponding to the last query is replied. The discovery load (or control overhead per node) is the average number of transacted discovery messages by each vnode during the querying period. Figs 18-a and 17-b depict the discovery load per node as the function of system size for different approaches. Figure 18-b demonstrates the results presented in Figure 18-a with better resolution (without Broad-Walk). As we can see in these figures, ElCore shows better performance compared to other solutions by generating minimal traffic transmission control overhead per node. This also proves that ElCore provides better scalability than others (in terms of discovery load per node for querying) as the system size increases.

Figure 18-c compares the average bandwidth consumption per node (during querying period) for different approaches and various system sizes. Bandwidth consumption per node indicates the amount of control traffic that each node generates during simulation time. The traffic load that we measured contains control, query and description messages that are transmitted between nodes. As we can expect, communication among RMC instances, to resolve various parallel resource requests, inevitably imposes unwanted traffic into the network. But, as we can see in Figure 18-c, the bandwidth consumption per node for our approach is significantly lower, compared to other approaches for all different system sizes. It can be also observed that, when we vary the system size from 5500 to 55000 resources, the bandwidth consumption for ElCore remains reasonable (with minimum changes), resulting in better scalability, compared to other approaches.

## 10. Conclusions

We have also presented a novel resource management architecture for large scale distributed computing environments. The proposed architecture contains a set of modules, which will dynamically be instantiated on the nodes in the distributed system on demand. Our approach is flexible to allocate the required set of resources for various types of processes/applications. It is a generic resource management approach, which can be applied to different large scale computing architectures and specifically it can be explored in Cloud systems. We must note that we have not specifically designed our approach for Cloud computing only. Instead, the proposed resource management architecture provides a generic solution (considering the general requirements of large scale computing environments), which can bring a set of interesting features (such as auto-scaling, multitenancy, multi-dimensional mapping, etc,.) that facilitate its easy adaptation to any distributed technology (such as Service Oriented Architecture (SOA), Grid and HPC many-core). Our resource management scheme can support auto-scale (the ability to provide budget to more-or-less expensive tasks) in two different aspects: dynamic resource allocation and dynamic module instantiation.

Using our approach, the mapping between processes and resources can be done with high level of accuracy which potentially leads to a significant enhancement in the overall system performance. System module instances would be automatically created when they are needed. Each module instance would be dependent on the type of application (e.g., real time, HPC, etc.), the computing resource (where the application starts to be executed), heterogeneity of resources, positioning of the resources in the system, network topology and interconnection between resources. Moreover, leveraging discovery components (RR-RPs) enables our resource management platform to dynamically find and allocate available resources that guarantee the QoS parameters on demand. We evaluated the performance of HARD and ElCore over highly heterogeneous, hierarchical and dynamic computing environments with respect to several scalability and efficiency aspects. Simulation results show significant scalability and efficiency of the proposed approaches over highly heterogeneous, hierarchical and dynamic manycore computing environments.

### Acknowledgements

The authors acknowledge the support of project FP7-ICT-2009.8.1, Grant Agreement No.248465, Service-oriented Operating Systems (2010-2013) [96-98] and of project Cloud Thinking (2013-2015), CENTRO-07-ST24-FEDER-002031.